\documentclass[conference]{IEEEtran}

\usepackage{titlesec}
\titlespacing{\section}{0pt}{6pt plus 2pt minus 2pt}{4pt plus 2pt minus 2pt}
\titlespacing{\subsection}{0pt}{6pt plus 2pt minus 2pt}{4pt plus 2pt minus 2pt}
\titlespacing{\subsubsection}{0pt}{3pt plus 2pt minus 2pt}{2pt plus 1pt minus 1pt}
\titlespacing{\paragraph}{0pt}{\parskip}{-\parskip}
\PassOptionsToPackage{usenames,dvipsnames}{xcolor}

\usepackage{amsmath}
\usepackage{amssymb}
\usepackage{physics}
\usepackage{dsfont}
\usepackage{mathrsfs}
\usepackage{wasysym}

\usepackage[hyphens,obeyspaces,spaces]{url}
\usepackage[utf8]{inputenc}
\usepackage{hyperref}
\usepackage{array, multirow}
\newcolumntype{P}[1]{>{\centering\arraybackslash}p{#1}}
\usepackage{tikz}
\usepackage{listings}
\usepackage{makecell}
\usepackage{subcaption}
\usepackage{flushend}
\usepackage{fancyhdr}
\usepackage{algorithm2e}
\usepackage{listings}
\usepackage{cprotect}
\usepackage{environ}
\usepackage{authblk}
\usepackage{amsmath}
\usepackage{mathtools}
\usepackage{svg}
\usepackage{xcolor}
\usepackage{graphicx}
\usepackage{MnSymbol,wasysym}
\usepackage{graphicx}
\usepackage{caption}
\renewcommand{\arraystretch}{1.25}
\usepackage{pifont}
\newcommand{\cmark}{\ding{51}} 
\newcommand{\xmark}{\ding{55}} 

\newcommand{\ignore}[1]{}

\newcolumntype{P}[1]{>{\centering\arraybackslash}p{#1}}
\usepackage{tikz}
\usepackage{amsmath}
\usepackage{amssymb}
\newcommand{\stt}[1]{{\small\path{#1}}}
\newcommand*\circled[1]{\tikz[baseline=(char.base)]{
            \node[shape=circle,draw,inner sep=1pt,semithick] (char) {#1};}}

\newcolumntype{H}{>{\setbox0=\hbox\bgroup}c<{\egroup}@{\hspace*{-\tabcolsep}}}

\newcolumntype{C}[1]{>{\centering\arraybackslash}p{#1}}

\begin{document}

\title{The CURE to Vulnerabilities in RPKI Validation}
\author{Donika Mirdita$^{\dagger\natural}$\qquad Haya Schulmann$^{\ddagger\natural}$\qquad Niklas Vogel$^{\ddagger\natural}$\qquad Michael Waidner$^{\dagger\S\natural}$\vspace{2mm}\\
{\small
$^\dagger$ Technische Universität Darmstadt \qquad $^\ddagger$ Goethe-Universität Frankfurt \qquad $^\S$ Fraunhofer SIT \qquad
$^\natural$ ATHENE} \\
}

\IEEEoverridecommandlockouts
\makeatletter\def\@IEEEpubidpullup{6.5\baselineskip}\makeatother
\IEEEpubid{\parbox{\columnwidth}{
    Network and Distributed System Security (NDSS) Symposium 2024\\
    26 February - 1 March 2024, San Diego, CA, USA\\
    ISBN 1-891562-93-2\\
    https://dx.doi.org/10.14722/ndss.2024.241093\\
    www.ndss-symposium.org
}
\hspace{\columnsep}\makebox[\columnwidth]{}}

\maketitle

\begin{abstract}
Over recent years, the Resource Public Key Infrastructure (RPKI) has seen increasing adoption, with now 37.8\% of the major networks filtering bogus BGP routes. Systems interact with the RPKI over Relying Party (RP) implementations that fetch RPKI objects and feed BGP routers with the validated prefix-ownership data. Consequently, any vulnerabilities or flaws within the RP software can substantially threaten the stability and security of Internet routing.\\
\indent We uncover severe flaws in all popular RP implementations, making them susceptible to path traversal attacks, remotely triggered crashes, and inherent inconsistencies, violating RPKI standards. We report a total of 18 vulnerabilities that can be exploited to downgrade RPKI validation in border routers or, worse, enable poisoning of the validation process, resulting in malicious prefixes being wrongfully validated and legitimate RPKI-covered prefixes failing validation. Furthermore, our research discloses inconsistencies in the validation process, with two popular implementations leaving 8149 prefixes unprotected from hijacks, 6405 of which belong to Amazon.\\
\indent While these findings are significant in their own right, our principal contribution lies in developing CURE, the first-of-its-kind system to systematically detect bugs, vulnerabilities, and RFC compliance issues in RP implementations via automated test generation. The statefulness of RPKI, the lack of rigorous RPKI specifications for recognizing bugs in the object suite, the complexity and diversity of RP implementations, and the inaccessibility of their critical functionalities render this a highly challenging research task. CURE is a powerful RPKI publication point emulator that enables easy and efficient fuzzing of complex RP validation pipelines. It is designed with a set of novel techniques, utilizing differential and stateful fuzzing. We generated over 600 million test cases and tested all popular RPs on them.\\ 
\indent Following our disclosure, the vendors already assigned CVEs
to the vulnerabilities we found. We are releasing our fuzzing system along with the CURE tool to enable the vendors improve the quality of RP implementations.
\end{abstract}

\section{Introduction}

The Border Gateway Protocol (BGP) is the defacto routing protocol of the Internet. Border routers use BGP packets to exchange information about the reachability of the prefixes allocated to their Autonomous Systems (ASes). A router can announce any prefix, including those it does not own. Routers that accept such bogus announcements route their traffic to the hijacking ASes instead of the target prefix \cite{bellovin1989security}. Such redirects can be exploited for blackholing and hijacking traffic \cite{china:telecom,ballani2007study,fb:out,u:tube,mitm:threat,indosat:hijack,turkey:hijack,vervier2015mind}.\\ 
\indent To validate ownership over announced prefixes the IETF standardized the Resource Public Key Infrastructure (RPKI). 
ASes first authenticate prefixes in RPKI: create a cryptographic key and bind their prefixes to AS numbers (ASNs) inside Route Origin Authorizations (ROAs). To validate BGP announcements and filter bogus routes and hence block hijacks, ASes enforce Route Origin Validation (ROV): compare received BGP messages against all ROAs and drop origins with unauthorized ASNs. 
RPKI also paves the way for prospective path validation mechanisms \cite{lepinski2017rfc,aspa,amir:ndss:2024}.\\ 
\indent The retrieval and processing of RPKI data is handled by Relying Party (RP) implementations, which periodically download and validate objects, like ROAs, from the RPKI and compile a list of all valid ASN-IP pairs inside a Validated ROA Payloads (VRPs) file. The border routers should retrieve the VRPs file and use it to validate all received BGP announcements. 
Border routers can only validate BGP announcements if the RP functions correctly and periodically provides data to the routers. If this continuous data stream stops, the validation data expires. As a result, the routers cannot validate RPKI and become vulnerable to BGP hijacks. Thus, availability and consistency are a major concern in RPs design.\\
\indent As of June 2023, 37.8\% of the networks utilize RPs to feed their BGP routers with RPKI data and validate BGP announcement with ROV \cite{DBLP:conf/uss/HlavacekSVW23,DBLP:conf/ccs/HlavacekJMSW22,shulman2022poster, rovista, apnicROV}. %
This increasing adoption of RPKI necessitates a deeper look into the correctness, stability, and consistency of RP software implementations. Ensuring these requirements of RPs is not trivial. The complexity of RPKI makes the creation of efficient, secure, bug-free, and RFC-compliant software implementations a complex task. Indeed, previous work found that due to missing limits on recursion depths, RP implementations were vulnerable to stalling attacks \cite{usenix-stalloris-21,DBLP:conf/sigcomm/HlavacekJMSW23,mirdita2022poster, koenvanhove}, which was fixed by introducing thresholds to all the RP implementations.
However, except for occasional bug reports, no systematic analysis of RPKI validation software has been performed to date. Adding to the problem, the community lacks a tool for generic, reliable, and continuous testing of RP implementations. The developers of RP software thus rely on manual testing and unit tests to ensure the correctness of their implementations. However, the existing unit test coverage, which the RPs provide, is not comprehensive enough to capture the RPs' complexity and cover all aspects requiring evaluation. Further, it is difficult for developers to compare the consistency of their validation process across the different RP implementations.\\
\indent Our goal is to analyze the current RP implementations, their resilience to errors, attacks and compliance with the standards. However, fuzzing RPKI is hard. As we show in this work, the statefulness of RPKI, the lack of rigorous specifications on a uniform processing logic for objects, and error recognition in the object suite, the complexity and diversity of RP implementations, coupled with the inaccessibility of their critical functionalities, renders this a highly challenging research task.\\
\indent To address the major challenges in fuzzing RPKI, we develop a new fuzzing system we call CURE for discovering vulnerabilities and bugs in arbitrary RP implementations. We use black-box fuzzing \cite{liang2018fuzzing} to fuzz RP implementations developed under different programming languages and models. Our tests include performing stateful fuzzing by mutating the objects we feed to the RPs \cite{ba2022stateful}. We combine the black-box fuzzing with manual code analysis and analysis of the RPKI RFCs and find major flaws in the widely used RP implementations. Detecting these issues is challenging not only because of the complexity of RPKI, but also because of the lack of rigorous definitions of a strict canonical behavior of the RPKI validation and the caching, in the specifications of the RPKI RFCs. We therefore also utilize differential fuzzing \cite{mckeeman1998differential} and use the inconsistency of the validated ROA payloads in caches to find vulnerabilities and bugs. We show that the inconsistencies, bugs, and vulnerabilities we find can lead to crashes and attacks against RPKI, such as downgrading RPKI validation or poisoning of RPKI trust anchors. In fact, an inconsistent behavior of RPs leads to discrepancies in decisions of different RP implementations regarding the validity of announced routes, even under benign network conditions.\\
\indent To support the automated fuzzing analysis, we develop a test system that includes the local RPKI hierarchy with an internal tree of Certificate Authorities and publication points with repositories that we generate for each test case. We optimize the performance of our system to enable 500K test cases per minute and support 13000 objects per second, which is faster than the fastest RP implementation Routinator, that processes up to 4545 objects per second.\\ 
\indent Our work shows that the maturity of RP implementations has not yet reached a level of strong security and consistency guarantees, despite the increased quality of RPKI objects and the rising rate of ROV enforcement in routers. The available RP implementations and tools lag behind the increasing utilization of RPKI in production environments. We hope our research will facilitate the improvement of the security and resilience of RP implementations. We make our CURE tool public and disclose the bugs we found to the RP developers.\\
\indent {\bf Contributions.} We make the following contributions:\\
\indent $\triangleright$ We develop the first automated system to analyze RP implementations, discovering novel vulnerabilities, and RFC violations. CURE presents a novel approach to testing complex validation pipelines. Instead of fuzzing isolated functions at a time, CURE executes the entire validation pipeline by efficiently emulating a complete repository structure.\\ 
\indent $\triangleright$ Combining manual code analysis with CURE, we perform the first comprehensive fuzzing of all popular RP software implementations. We find a total of 18 previously unknown critical vulnerabilities, including DoS and poisoning of VRP cache with 7 additional RFC inconsistencies leading to differing validation results in RP implementations. Conceptually, we show that the complexity of the validation pipeline, the vagueness of RFCs, and the immaturity of implementations lead to vulnerabilities and diverging validation results. \\
\indent $\triangleright$ Using CURE, we perform an analysis of global RPKI data, finding inconsistent processing results across all RP implementations. We show that the inconsistencies apply to prefixes of large network operators and lead to discrepancies in validation results.\\
\indent $\triangleright$ We make CURE open for public, to help improve RP software quality and research into the area of RP security. CURE provides developers with comprehensive and novel functionality, lacking in available tools, like fuzzing the entire validation pipeline, constructing arbitrary repository structures, and manually testing arbitrary test cases against any RP.\\ 
\indent {\bf Ethics and disclosure.} Our research was conducted on our isolated test environments. Thus, all crashes, bugs, and vulnerabilities we found were not propagated to the Internet. In our measurement we follow the ethical guidelines for network measurements \cite{durumeric2013zmap,partridge2016ethical}. 
We notified the affected vendors, and five CVEs based on our work have already been registered: CVE-2023-39914, CVE-2023-39915, CVE-2023-39916, CVE-2022-3029, CVE-2022-3616.\\
\indent {\bf Organization.} Section \ref{sc:works} provides an overview of related research and Section \ref{sc:overview} overview of RPKI. Section \ref{sc:hurdles} discusses obstacles for fuzzing the RPKI. Section \ref{sc:cure} introduces CURE. 
Section \ref{taxonomy} describes the vulnerabilities and inconsistencies we found in RPs. Section \ref{sc:impact} evaluates the impact of the discovered vulnerabilities on the Internet and quantifies the affected networks. We conclude in Section \ref{sc:conclusions}.

\section{Related Work}\label{sc:works}
Fuzzing is one of the most popular and efficient software testing mechanisms available. It can be tailored to a specific target and has systematically helped developers improve their software and identify vulnerabilities \cite{DBLP:conf/uss/KruppGR22}. Fuzzing is flexible and accommodates a variety of targets: OS kernels \cite{schumilo2017kafl}, protocols like TLS \cite{de2015protocol}, and user applications \cite{DBLP:conf/uss/KandeCPJSTR22}. Considerable work has been done in the development of fuzzing frameworks. PeachFuzzer \cite{eddington2011peach} is a powerful framework that enables mutation and generation-based fuzzing, ultimately acquired by Gitlab in 2020. AFL, and its successor AFL++ \cite{AFLplusplus-Woot20}, are some of the popular frameworks for application-based fuzzing. Google created OSS \cite{serebryany2017oss}, a comprehensive software suite to fuzz open source codebases, which employs a toolkit made of multiple fuzzing algorithms. These tools are powerful in finding vulnerabilities, but they are limited in their applicability; as fuzzing requires inputting the test objects into the investigated binary, developers need to create a test harness around functions they want to fuzz. This workflow requires major manual work for each function and is traditionally limited to a single test object per test run. \\
\indent {\bf Fuzzing Internet protocols.} Fuzzing has been used to study core Internet protocols, such as DNS and BGP \cite{kakarla2022scale,wang2013rpfuzzer}. 
As we show in this work, fuzzing RPKI is fundamentally different from previous work on fuzzing Internet protocols and requires a novel viewpoint on testing complex applications. First, RPKI involves the validation of cryptographic digital signatures on tested objects. Automatically generating random, yet validly signed, input objects with a fuzzer is challenging since signatures depend on the object content and thus need to be re-generated for every object created by the fuzzer. Second, generating meaningful RPKI objects is hard due to their complex structure and dependencies. Third, stateful fuzzing is a hard problem in network fuzzing due to the computational complexity involved \cite{ba2022stateful}, and RPKI is stateful since RPs cache the validated objects. Fourth, fuzzing the RPKI validation requires creating a custom PP that supports the required properties for automatically testing the behavior of RPs on different arbitrary input objects. The need for a custom PP arises from the validation pipeline of the RPs, which requires several validation steps to succeed, involving multiple interdependent objects and signature validations before the testing object is even parsed. The PP setup is unique for every object, necessitating a flexible PP that can efficiently create a valid repository around each test-object. Requiring multiple additional objects tailored at every test case makes fuzzing the RPKI more complex and computationally intensive than fuzzing BGP/DNS packet processing. Therefore, in this work, we detach from the idea of fuzzing as ``testing a single object per iteration''. Instead, we create a custom tool that efficiently generates a complex setup for every test case consisting of multiple RPKI objects. Despite the complex setup, our tool can test thousands of objects per second. We next discuss conceptional fuzzing approaches and explain our approach.\\
\indent {\bf Fuzzing approaches.} There are multiple possible fuzzing strategies that could be applied to RP testing. White-box fuzzing \cite{godefroid2008automated,godefroid2008grammar} takes source code and software paths into account, thereby allowing tailored inputs to maximize code coverage, usually through symbolic execution and feedback loops that the test case generator can use to find new unbeaten paths. Black-box fuzzing \cite{liang2018fuzzing} is the opposite of whitebox. In black-box fuzzing, the fuzzer receives no feedback from the target software, only being able to observe the results of the process and adapt according to the output. Gray-box fuzzing \cite{bohme2016coverage} lies somewhere between whitebox and blackbox. The fuzzer is privy only to a subset of execution information, but has more insights into target behavior than blackbox setups. Grey-box fuzzers usually retrieve execution information through language-specific compiler setups or additional tooling and use gained information to run coverage-based fuzzing \cite{bohme2016coverage,yan2020pathafl}. Coverage-based fuzzing optimizes test objects by taking execution paths in the target binary into account. A major drawback of coverage-based fuzzing is the potential for path explosion, an inherent issue with large and complex codebases. The complexity of the RPKI processing pipeline, including a multitude of field and signature validations, thus limits the applicability of coverage-based fuzzing
In this work, we utilize a blackbox fuzzing approach for testing RPs without additional language-specific tooling. We designed our middleware for RPKI fuzzing to simultaneously test any RP, open source or proprietary, without requiring any adaption of binary compilation or language-specific tooling. We also combine the blackbox fuzzing with analysis of RFCs and code analysis of RPKI implementations to understand and identify coding errors, and trace them to core specification problems.\\
\indent {\bf Mutation-based object generation.} Fuzzing requires an efficient generation of test objects to feed objects to the tested binary. Mutation-based object generation is one of the most common test generation processes for fuzzing. This approach utilizes a corpus of valid objects manipulated through bit flips, random cuts, deletions, or character insertions. The approach generates diverse mutated files to test all potential parsing and validation corner cases in the target code. This method is particularly suitable for testing parsing modules of binaries as it can create malformed entities similar to well-structured objects. Mutations can be generated through generic mix\&match and random mutations \cite{afl,brubaker2014using} or genetic algorithms \cite{zhou2020web} that take advantage of feedback loops in grey- or white-box fuzzing. Mutation-generated test cases are useful in efficiently creating a large test corpus to trigger different execution paths. However, it does not take the correct structure of objects and their encoding into account and is thus very limited in moving past the parsing routines of the target. \\
\indent {\bf Structure-aware fuzzing.} Improving on this limitation, structure-aware fuzzing, also referred to as grammar-based fuzzing, is a test case generation mechanism that enables the creation of objects that follow strict formatting and encoding requirements according to the target software specifications \cite{wang2019superion,torres2020nfdfuzz,kim2022efficient}. 
The advantage of using structure-aware fuzzing is its ability to move past the object parsing stage by providing correctly encoded and formatted objects, allowing testing of deeper parts of binary execution. We use this approach to create manipulated but valid complex RPKI objects to test for security issues in parts of the code beyond object parsing. With structure-aware fuzzing, we can stress test the internal processing logic of RPs, focusing on the validation pipeline. This ensures our test cases do not get stuck at the first few parsing checks but lead the way to cache inconsistencies. \\
\indent Recognizing the advantages of random mutations and structure-aware fuzzing, we dissect the RPKI software suite using multiple approaches. 
We create an object generation tool based on random mutations and feed it to the RPs over our CURE tool to test the complex RPKI object parsing of the RPs. Additionally, we move past the object parsing with a second, structure-aware object generation approach capable of producing well-formatted RPKI objects, with specific random mutations within its fields. These well-formatted objects are also fed to the RPs via CURE. 
Additionally, we employ code analysis of the RP software and a manual analysis of the RFC requirements for the validation process. For comprehensive testing on the RPKI, we isolate requirements in the standards to design test cases that target them.

\section{\hspace{1mm}Background on RPKI Validation Process}\label{sc:overview}
\begin{figure}[t!]
    \centering
    \includegraphics[width=0.7\columnwidth]{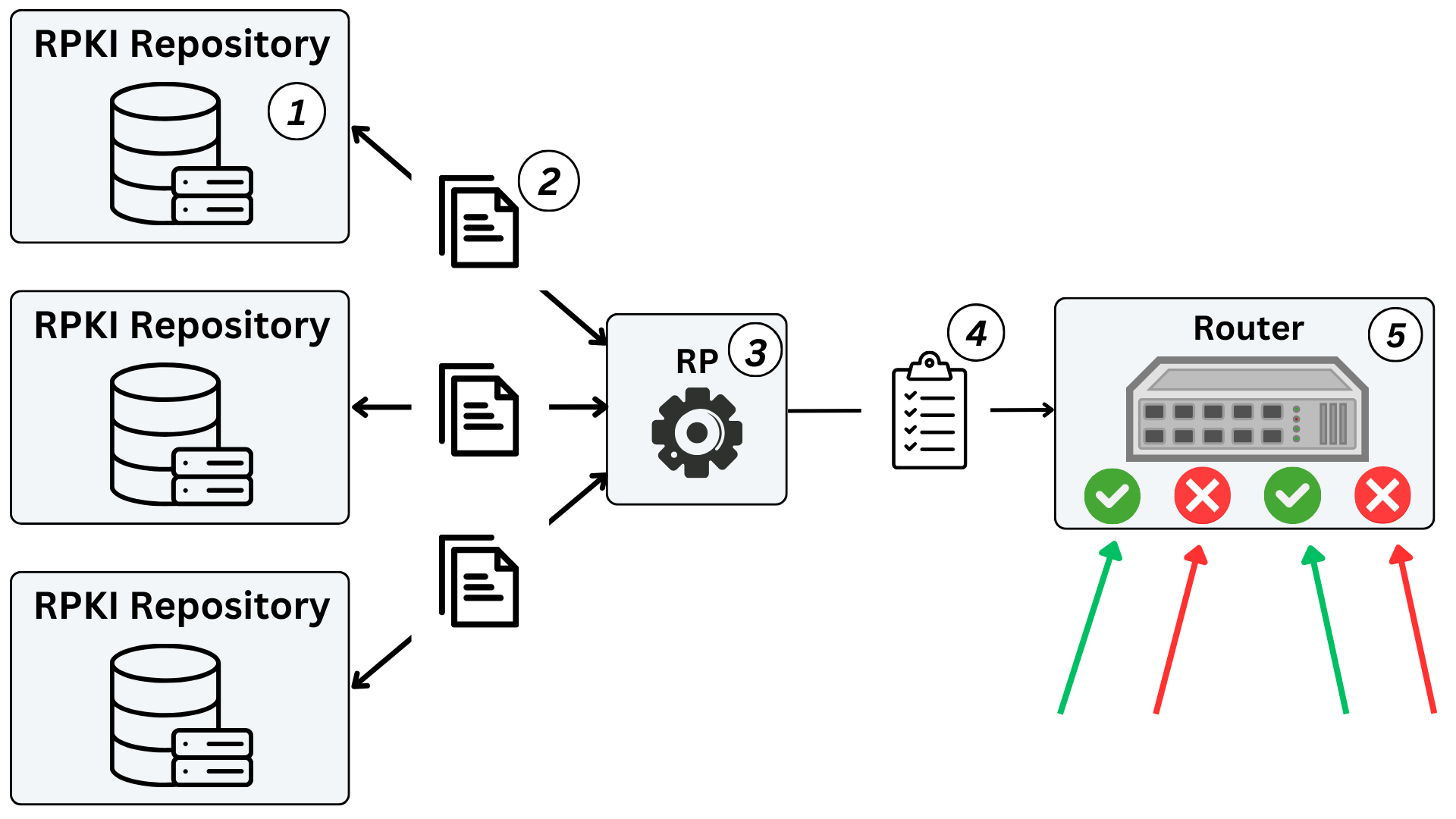}
    \vspace{-5pt}
    \caption{Overview RPKI.}
    \label{fig:rpki_fig}
    \vspace{-15pt}
\end{figure}

RPKI aims to provide authenticated information on prefix ownership, which border routers can use to make secure routing decisions in BGP. An overview on the RPKI ecosystem is given in Figure \ref{fig:rpki_fig}. RPKI information is stored in repositories (1) hosted on distributed hierarchically organized Publication Points (PPs) [RFC6481]. Repositories are operated either in hosted mode, e.g., by the Regional Internet Registries (RIRs), or in delegated mode by the resource owner. \\
\indent Repositories store cryptographically signed Route Origin Authorizations that detail which AS is authorized to announce a particular prefix over BGP. The RPKI also contains additional files like certificates, Certificate Revocation Lists (CRLs), Manifests (MFTs), Autonomous System Provider Authorization (ASPA), BGPsec certificates, and GhostBuster Records (GBR). Additional files contain additional information for validation different aspects of BGP announcements, or they ensure the integrity and authenticity of the RPKI infrastructure. An interested reader is referred to Appendix \ref{ap.sc.objects} for an extended explanation of the RPKI objects. All objects in RPKI, including ROAs, are signed by the private key of the Certificate Authority (CA) that manages the resources detailed in the object. A CA usually corresponds to a single business entity owning one or multiple ASes. 
CAs can be run by Regional Internet Registries (RIRs), or by resource owning
clients, e.g., the Local Internet Registries (LIRs).\\
\indent {\bf Relying Party software.} In RPKI, routers do not directly interact with repositories. Instead, the download and validation of objects is performed by RP implementations, as shown in Figure \ref{fig:rpki_fig} \circled{3}. Generally, each system that wants to utilize the RPKI installs its own RP, which continuously retrieves and validates objects from all RPKI repositories available on the Internet by establishing a chain of trust and delegation stemming from the root trust anchors, the five RIRs. The workflow of RPs for downloading and validation consists of multiple functions: recursive data fetching from PPs discovered over URIs in CA certificates, validating RPKI objects, compiling valid prefix-ASN pairs in a single VRPs file Figure \ref{fig:rpki_fig} \circled{4}, and finally exposing this file for download to BGP routers of the system. After the RP finishes one iteration, the VRPs file is downloaded by border routers to perform Route Origin Validation and filter invalid BGP routes, Figure \ref{fig:rpki_fig} \circled{5}.\\
\indent {\bf Publication Points traversal.} Bootstrapping the validation process of RPs is achieved over hardcoded Trust Anchor Locators (TALs) files [RFC6490], included in each RP installation. In a default installation, each RP contains five TALs containing HTTPS URIs to the root CA certificates of the five global RIRs. These RIR certificates are the root of trust in the RPKI and are inherently trusted. The RPs download and parse the root certificates. Each certificate contains a mandatory extension named Subject Information Access (SIA), which lists the URI of the repository that contains the objects signed by this certificate. The RPs follow the URI and download all objects from the repository, then use the public key of the CA certificate to validate them. After download and validation, all objects are stored in a local cache, using the unique URI of each object as its local file path. If the RP encounters new CA certificates stored in the repository it downloaded, it again checks each certificate's SIA extension to enumerate additional RPKI repositories' locations. Unknown repositories are added to a queue and iteratively downloaded and validated. With this workflow, the RPs traverse all public RPKI repositories on the Internet, downloading and validating all objects.\\ 
RPKI supports two protocols for fetching objects from PPs from the URI fetched from the SIA extension of a certificate, shown in Figure \ref{fig:rpki_fig} \circled{2}. RRDP is the primary protocol to retrieve RPKI objects [RFC8182] and is the default protocol in all common RP implementations. In case of failures in RRDP, RPs fall back to the legacy rsync protocol, a simple file synchronization protocol that identifies updates to files over their hash values. 
In contrast to rsync, RRDP is more sophisticated, providing additional logic for efficient incremental updates of local RP caches. It is based on HTTPS and uses three additional types of objects \textit{notification.xml}, \textit{snapshot.xml}, and
\textit{delta.xml} to synchronize the content between the PP and the RP cache. The entry to each PP is identified by the notification file, located at the URI in the SIA extension of the CA certificate. The notification file contains a serial number, allowing the RP to quickly identify if the repository content changed since the last validation run. It further contains URIs and hashes of a file containing the entire repository content, the snapshot file, and an arbitrary number of files containing incremental updates from serial numbers, referred to as delta files. 
During the first encounter of a PP, the RPs always download the snapshot file to retrieve and store the entire repository content in their local cache. In following validation runs, the RP can restrain from downloading the entire repository and instead only download the delta files spanning from the serial number of its local cache to the current state of the PP, i.e. the current serial number of the notification file. In case the processing of delta files encounters an error, the RP falls back to downloading the entire repository content from the snapshot. To ensure the integrity, the snapshot and all delta files also contain the same session\_id and the serial\_number as the notification file.\\
\begin{figure}[t!]
    \centering
    \includegraphics[width=0.9\columnwidth]{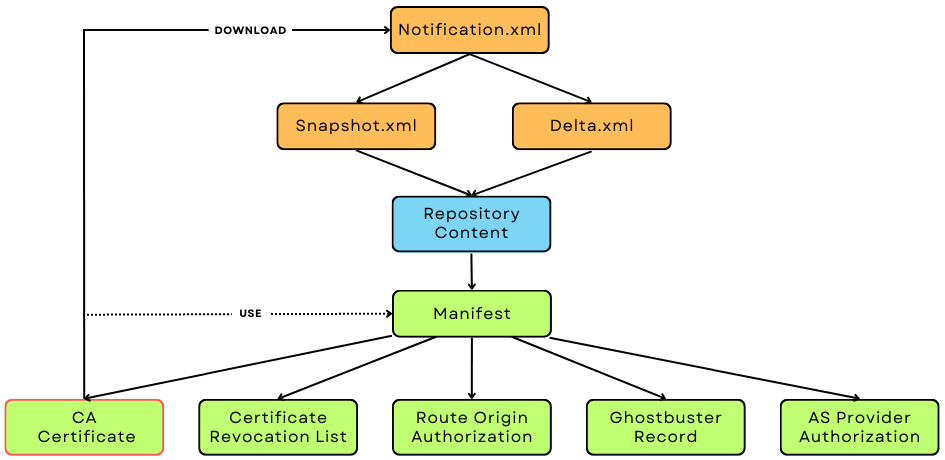}
    \vspace{-5pt}
    \caption{RPKI objects download flowchart.}
    \label{fig:rp_validation}
    \vspace{-15pt}
\end{figure}
\indent {\bf RPKI validation process.} After downloading the snapshot or applying all deltas, the RP follows a strict processing order to validate retrieved objects, visualized in Figure \ref{fig:rp_validation}. For each repository, the CA certificate is validated before any content is parsed. RPs start the validation process of CA certificates by downloading the five RIR certificates. All child certificates of a CA can be found in its RPKI repository, in the case of the root CAs the root repositories. After validating all certificates found in the root repositories of the RIRs with the trusted public keys, the RPs traverse the tree, following the SIA extensions of the now trusted child CA certificates to discover more certificates in their respective repositories, which are validated by their respective parent. By iterating down the CA tree, this process establishes a trust chain from the five root nodes to each CA certificate, following the issuing of resources. 
Every time the CA certificate of a given repository is validated, the RPs can use its public key to also validate all RPKI objects in the repository of this CA. Looking at the validation of repository content, we must distinguish between CRLs and all other RPKI objects. CRL validation is straightforward; the RP validates the signature on the CRL with the public key of its CA certificate and confirms that the CRL formatting complies with the requirements in [RFC5280]. For all other RPKI objects, including ROAs, each file contains two signatures, complicating validation. The object's payload, like the ROA content, is signed with a one-off key pair, i.e., a key only used for a single signature. The generated one-off public key is contained in the ROA file inside a X.509 certificate to allow validation of the ROA content. This object-specific one-off certificate, referred to as the End-Entity (EE) certificate, is signed by the CA. Thus, the RP first validates the signature of the content with the one-off public key in the EE-certificate, then validates the signature on the EE-certificate with the CA certificate. Additionally to the described signature validation, RPKI also includes integrity validation over Manifest files [RFC6486].
Each repository should contain a single Manifest file that lists all the objects currently in use by the CA, identified over their storage URIs and hashes. To validate the integrity of a repository, the RP compares the manifest content against all files downloaded from a PP, identifying missing or extraneous files (e.g., due to attack or misconfigurations). After ensuring integrity, the RP processes all files listed in the manifest, validating their signatures and checking formatting. Following processing, each file is stored in the local cache. This processing is repeated for each discovered PP until all discovered files finish processing and are added to the validated cache. Finally, the RP compiles the content of all discovered ROAs and outputs them into the VRPs file for the routers. After outputting the VRPs, the RP goes idle until the next validation run starts in a pre-defined interval.  
\section{\hspace{1mm} Resolving Hurdles Towards Fuzzing RPKI}\label{sc:hurdles}
Testing RPs is a hard and complex task. RPs do not run in isolation, they integrate within an RPKI environment to fetch and process cryptographic objects. Therefore feeding each RP with files to test requires the creation of a customized environment with trust anchors, certificates, and other auxiliary objects. Further, RPs are stateful, caching validated RPKI objects for consecutive validation runs inside a locally validated cache. Stateful fuzzing was shown to be a hard problem due to the computational complexity involved \cite{ba2022stateful}. Also, the diverse landscape of programming languages of the RP implementations further complicates the manual effort of white- or greybox RP fuzzing. Each RP requires individual effort and unique setups. In this section we explain why existing fuzzing methods are not suitable for fuzzing RPKI validation with RPs, clarifying what hurdles we resolved while developing our CURE fuzzing tool.\\
\indent {\bf Complexity of function decoupling.} Currently available RP software is characterized by tight coupling of components and high code complexity. This makes fuzzing RPKI implementations hard. Traditional fuzzing requires isolating functions of interest, extracting them from the code and adapting their interface to accept manipulated inputs generated by the fuzzer. When implementations are tightly coupled and partially state dependent, it creates problems with the scalability and comprehensiveness of the fuzzed functions. While it is possible to fuzz specific routines, it is labor-intensive to build a testing harness for all functions of interest in all RP implementations, and create all objects required as input in the appropriate format. The complex interaction between RPKI objects in the validation pipeline makes comprehensive fuzzing of the validation process extremely difficult with traditional fuzzers.\\ 
\indent {\bf Manual code adaptation.} Some components of the implementations are not externally accessible, thus fuzzing them requires adaptation of the source code or a full extraction of the function including child functions, further increasing manual effort. Fuzzing all steps in the validation process, including all interactions with other RPKI objects, is thus immensely time-consuming. Repeating the setup process for each RP and after each major change of the constantly evolving RP software is not scalable. Thus, whitebox and greybox fuzzing can only be applied to small parts of the validation process, but they are not suitable to test the entire validation pipeline.\\
\indent {\bf Limitations of existing tools.} RPKI publication point software is built to be rigid and highly deterministic. The room of human error is minimized by automating the publishing pipeline away from the user and introducing additional checks to ensure high reliability in published objects. While the restrictive operation of common software implementations is important to ensure the high quality of published RPKI objects, it severely limits the flexibility given to the user; a flexibility that is required for stress testing the RPs with malicious and malformed objects. For default configurations, users neither create nor interact with any RPKI objects. Instead, they provide data to the PP, which then creates the corresponding objects. For example, users do not provide a complete ROA object to the publication point. Instead, a user inputs the data required for the creation of the ROA, i.e., IP-prefixes and an ASN. The publication point uses the data to create a ROA object and generate all required auxiliary objects. The common implementations of publication points are thus not well suited as a starting point for the implementation of a fuzzing tool, lacking the flexibility and speed to test arbitrary objects in a short amount of time.\\
\begin{table}[t!]
\scriptsize
\renewcommand{\arraystretch}{0.7}
\centering
\begin{tabular}{l|c|c}
\centering{\textbf{Feature}} & \textbf{Krill} & \textbf{Fuzzing} \\ \hline
\textbf{Input} & Parameters & Objects \\ 
\textbf{Pipeline} & Strict & Dynamic \\
\textbf{Key Security} & Secure & Not relevant \\
\textbf{Max. Speed} & 10 obj/s & 1000s obj/s \\
\textbf{Formatting} & Checked & Unchecked \\
\textbf{RP Control} & No & Yes 
\end{tabular}
\caption{Krill RPKI PP vs. the requirements for fuzzing.}
\vspace{-15pt}
\label{tab:table_pp}
\end{table}
\indent {\bf Resolving the limitations.} To resolve these limitations in existing fuzzers, we develop a language-agnostic, black-box, continuous testing middleware for evaluation of RP implementations. Our novel fuzzing implementation emulates a highly flexible RPKI publication point, built on-top of existing RPKI libraries. We remove features not required for fuzzing, while introducing flexibility and efficiency improvements. Our tool emulates all the core features of a PP while leaving the object generation open and flexible for the user to test RPs against fuzzed inputs. Further, our custom PP provides a set of additional features and characteristics not provided by traditional implementations, which we explain in Section \ref{ssc:cure}. A key insight motivating our design is that we do not define our implementation as a rigid pipeline aimed at publishing ROAs and maintaining a valid repository. Instead, the tool should "\textit{repository-fy}" any RPKI object, i.e., it takes an arbitrary RPKI object as input and builds a scaffold repository around this object. The scaffolding does not follow a fixed pipeline but is instead flexible, adapting dynamically to which type of RPKI object is input into the tool. This functionality is not provided by any existing publication point software.\\
\indent {\bf Eliminating false positives.} Each RP is treated as a blackbox by our fuzzer, and conclusions about internal problems are derived based on the output of the tested RP. Applying the generated fuzzing test cases to the entire validation process instead of individual functions, enables us to be immune to false-positive crashes. The RPs in our setup operate identically to real-world deployments. Each input that crashes an RP directly translates to a vulnerability as the fuzzing framework emulates the normal operation of RPs. This is not true for other fuzzers. They might find a crash in an extracted function, but this crash might not lead to a real vulnerability, either because the malformed input is escaped before reaching the investigated section in the code, or because the software containing the fuzzed function implements proper error handling. \\
\indent In summary, an environment for running a comprehensive vulnerability analysis of the RPs needs to provide features not supported by existing RPKI software and libraries; see Table \ref{tab:table_pp}. The available tools do not offer the features for realizing a functional and fast modern fuzzer. Thus, we introduce our solution for fuzzing the RP implementations.

\section{Developing the CURE}\label{sc:cure}

\begin{figure}[t!]
    \centering
    \includegraphics[width=0.47\textwidth]{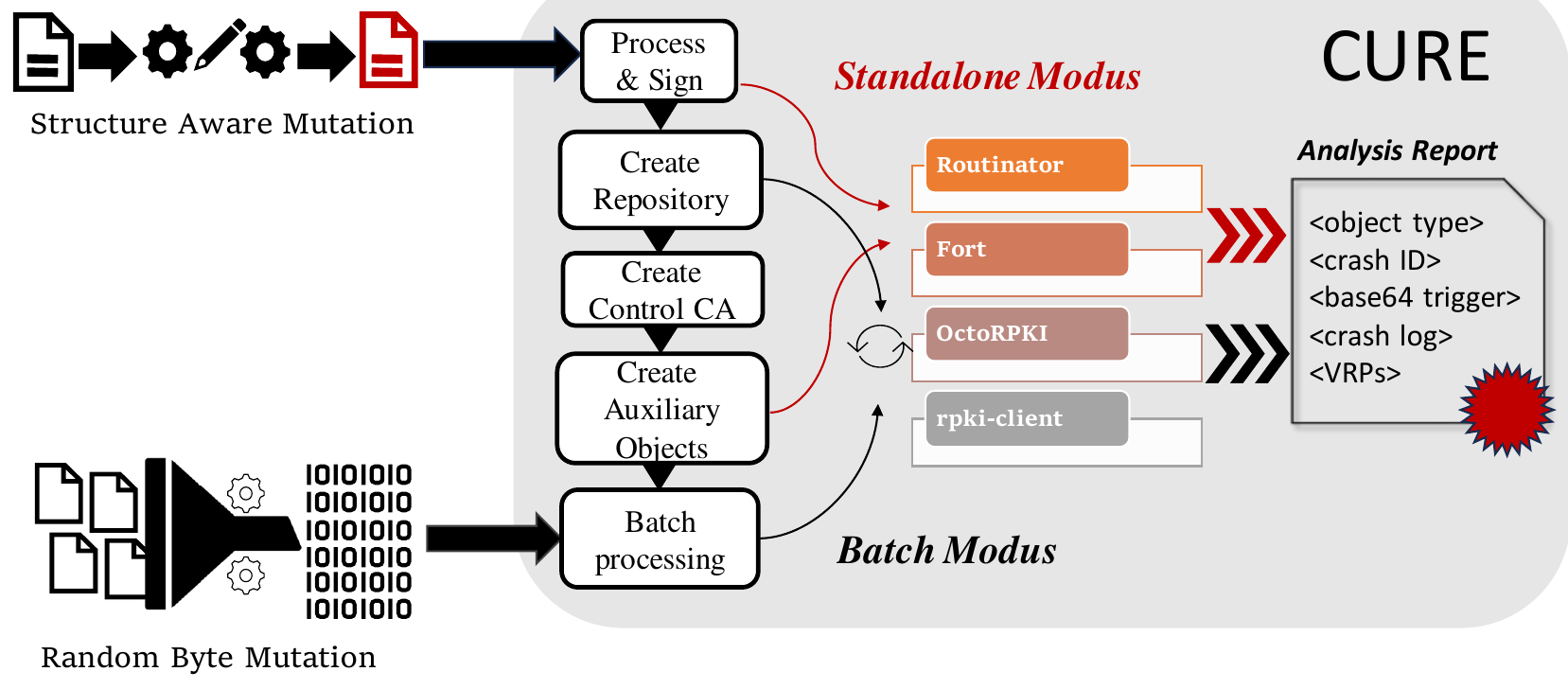}
    \vspace{-5pt}
    \caption{CURE fuzzing framework.}
    \label{fig:2mode-pipeline}
    \vspace{-20pt}
\end{figure}

In this section, we present our framework for fuzzing-based black-box vulnerability analysis of RP software. The core component of our setup is the Comprehensively Usable RP Evaluator (CURE) middleware we have built. It emulates a PP repository while monitoring the blackbox execution of RPs and performs differential analysis on RP output and states. CURE feeds the RPs with test cases using randomized mutations and structure-aware object generation. An overview of the fuzzing framework is illustrated in Figure \ref{fig:2mode-pipeline}.\\
\indent To create test cases, we use two approaches that generate a large corpus of malicious or malformed RPKI objects for fuzzing. Object creation is efficient and fast, generating thousands of objects per second. 
Feeding the generated RPKI objects to the RPs is done over CURE. It interfaces with the corpus generation, processes the generated objects and inserts them into an RPKI repository, which is then exposed to the RPs over one of the RPKI data exchange protocols. The RPs are managed by the RP interaction module, which handles execution of the RPs as well as crash detection. The module additionally detects anomalies in the RP caches and creates a human-readable report, detailing all the detected inconsistencies and vulnerabilities. \\

\begin{figure}[h!]
    \centering
    \includegraphics[width=0.85\columnwidth]{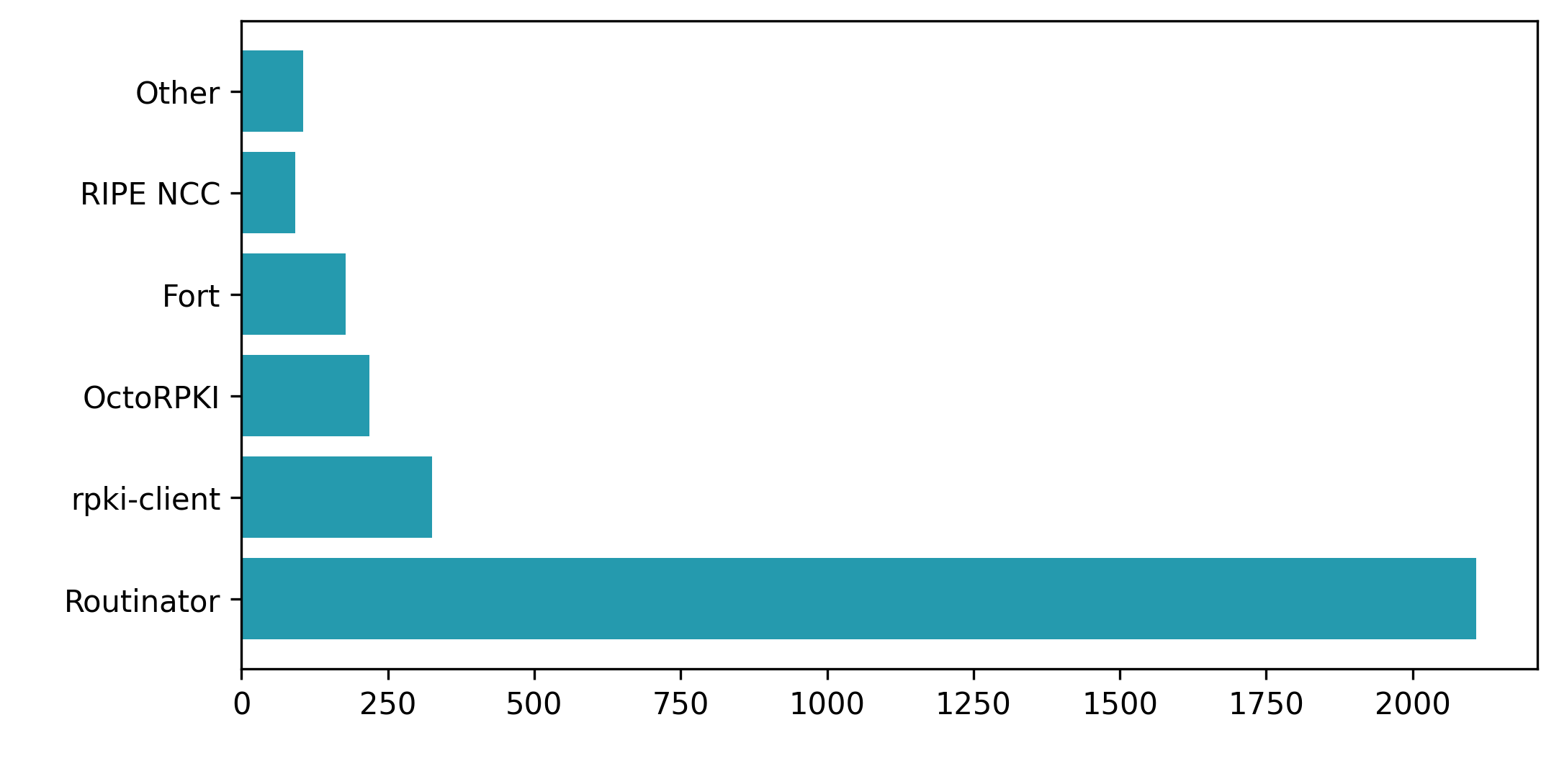}
    \vspace{-10pt}
    \caption{RP implementations popularity distribution, April 2023.}
    \label{fig:rp-usage}
\vspace{-20pt}
\end{figure}

\indent We run our blackbox analysis against Routinator, FORT, OctoRPKI and rpki-client as the four major RP implementations currently in use on the Internet according to our measurements, see Figure \ref{fig:rp-usage}. RIPE NCC Validator, whose support and maintenance has been discontinued since 2021, was excluded from our study.

\subsection{Generating RPKI Objects}
Detecting vulnerabilities and inconsistencies is based on feeding manipulated objects to the RPs and probing how their processing handles unexpected inputs. Generally, RPs digest three data types and a total of nine different schemas. For object generation, we identified the data types (XMLs, ASN.1 and vCard), and compiled the schema for each file type; see Table \ref{tab:rpki_objects}. 
\begin{table}[t!]
\scriptsize
\renewcommand{\arraystretch}{0.7}
\centering
\begin{tabular}{ p{1.2cm}|C{0.5cm}|C{0.5cm}|C{0.5cm}}
& \textbf{ASN.1} &  \textbf{XML} & \textbf{vCard}  \\
\hline
\textbf{roa} & X & &\\
\textbf{crl} & X& &\\
\textbf{mft} & X &  &\\
\textbf{certificate} & X&  &\\
\textbf{aspa} & X&  &\\
\textbf{gbr} &  &&X\\
\textbf{snapshot} && X &  \\
\textbf{delta} && X & \\
\textbf{notification} && X & 
\end{tabular}
\vspace{-5pt}
\caption{Supported RPKI Objects and Data Structure.}
\label{tab:rpki_objects}
\vspace{-15pt}
\end{table}
Our object generation strategy accommodates all RPKI object schemas and types. 
RPKI objects and their processing is characterized by strict formatting and encoding requirements. We thus approach the object generation two-ways. To target parsing and formatting errors, we utilize random byte mutations as unexpected input, non-conforming to file schemas. To reach deeper into the internal validation logic, we implement structure-aware RPKI object generation. We run our algorithms on all data formats, with our generation process able to create millions of objects in a 24h period.\\
\indent \textbf{Random byte mutation.} The first object generation strategy we employ is random mutations. This approach does not preserve the strict structure of objects and thus primarily targets programming, parsing and schematic errors. Our algorithm applies random byte mutations to a base corpus of RPKI objects, inserting, deleting and changing the order of bits inside the objects.\\
\indent \textbf{Structure aware generation.} We use structure-aware mutations to create schema-abiding and correctly encoded test cases to ensure that parsing checks are successfully passed and therefore test the internal logic of RP processing. To mass generate our objects we use \cite{atheris}, a fuzzing framework capable of providing structure-aware mutations on predefined file formats.

\subsection{CURE Architecture}\label{ssc:cure}
\begin{figure}[t!]
    \centering
    \includegraphics[width=0.9\columnwidth]{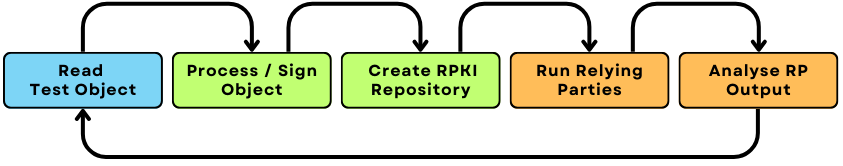}
    \vspace{-5pt}
    \caption{Finite state machine of CURE.}
    \label{fig:fuzzer_pipeline}
    \vspace{-10pt}
\end{figure}

To address the limitations of available RPKI tools in their applicability for fuzzing RP software, we introduce CURE for RP testing. CURE provides the same core functionality as a comprehensive RPKI PP, i.e., it handles keys, creates and manages a tree of CAs, signs objects, generates files to interact with the RPs and exposes these files over HTTPs. From the perspective of the RPs, CURE looks and acts like a standard PP. However, CURE has the flexibility to scaffold a complete and valid repository around arbitrary RPKI objects. Further, it can serve a fuzzing corpus. Speed is also a major concern in the operation of CURE as it needs to maximize processed objects and optimize the rate at which inputs can be tested. Our tool also provides an interface to directly interact with RPs, starting their validation and extracting validation results and crashes. We developed this interaction to be language agnostic, i.e., supporting RPs written in any arbitrary programming language. An overview of CURE is shown in Figure \ref{fig:fuzzer_pipeline}. \\
\indent The tool implements the entire pipeline to publish RPKI objects of any type. Objects are read, signed and then embedded in a valid RPKI repository. CURE then executes the connected RPs, which download and process the repository. After execution, CURE analyzes the RP output and logs any problems found, including crashes and cache inconsistencies.\\
\indent Our goal with CURE is to provide an easy to use tool for the RPKI community to test and improve the resilience of RP implementations. We therefore design it to be built loosely coupled and easy to extend, allowing the community to extend CURE for their unique requirements. This includes extending the tool to cover new RP implementations and new RPKI objects.
In the following, the core components of CURE are introduced, emphasizing the challenges and problems that motivated our design. 

\subsection{Object Generation Interface}
CURE serves as the middleware to distribute generated manipulated objects to the RPs. It thus needs to interface with the object generation algorithm. CURE supports two different modes of operation. \\
\indent \textbf{Batch-mode} allows parallel testing of a large corpus of objects. It is optimized to run continuously, consuming objects from a supplied folder and feeding them to the RPs in batches of a configurable size. Interfacing with batch-mode is straightforward, the user provides a folder URI containing RPKI objects and supplies the object type and desired batch size. CURE then feeds these objects to the RPs. Batch-mode allows for continuous operation, i.e., it idles if no new objects are present in the folder and continuous processing once new objects are inserted. This allows CURE to run independent of the test case generation process. While the generation continuously feeds new objects into the folder, CURE consumes them and distributes them to RPs. Batch mode implements the fuzzer capability of running indefinitely, testing new objects and searching for vulnerabilities until the fuzzer is stopped. \\
\indent \textbf{Standalone-mode} allows single-object testing over the command line. CURE takes a single RPKI object as input and creates a valid RPKI repository around it, including auxiliaries. It then executes the validation of all connected RPs and analyses their logs. Finally, it returns a report to the user, containing error analytics. Standalone mode is targeted and optimized for a single validation run of the RPs and does not include multiprocessing capabilities, since the overhead of spawning and interacting with created processes results in a slowdown of execution speed in a single run.

\subsection{Building a Repository}
The complexity of the RPKI repository structure complicates the distribution of the objects. The objects need to be incorporated into a valid RPKI repository to ensure RPs process the file. Further, most objects have inter-dependencies, e.g., a manifest needs to link to its issuer inside a specific field of the object, and must contain the names and hashes of all other files in the repository. Thus, objects cannot be merely inserted into a template repository and then distributed to the RPs. Instead, CURE needs to read the object, {\em dynamically} adapt fields and values inside the object, and then create the required auxiliaries for the repository to distribute to the RPs. To test a single ROA, CURE needs to create 8 auxiliary RPKI objects, requiring generation of 5 signatures, 3 hashes and adapting a multitude of fields inside created files. CURE achieves this as follows: 
\subsubsection{Creating a Repository and Processing Objects} In a first step, CURE creates an RPKI repository, by generating a root CA certificate, a TAL that points to the certificate, a manifest file and a CRL. This step is uniform and independent of the object type.
Following initial repository creation, the external test object is processed. The processing heavily depends on the object type, as each object type has individual fields that need to be handled.\\
\indent {\bf Dependencies:}\\
\indent $\bullet$ 
For ROAs, ASPAs, and GBRs, the object content is self-contained, no field inside the object itself has to be adapted. The content does not contain any pointers to other objects in the repository. Still, an End-Entity (EE)-certificate signed by a CURE certificate authority needs to be appended to the objects to publish them. For this, CURE uses the created root certificate to generate a child CA that signs the respective object and appends an EE-certificate to the content. Since the EE-certificate contains values that are directly related to the content of the object, like its digest and a signature, it cannot be precomputed and has to be generated dynamically for each individual test case.\\
\indent 
$\bullet$ For certificates and CRLs, the processing requires more effort as they contain dynamic fields that directly relate to repository content.  
A certificate or CRL generated by an object generation algorithm from a template does not have correct values in externally dependent fields and is not validly signed. Since the signature depends on the content, it cannot be hardcoded, but has to be computed with the private key of the CA that issues the object {\em after the object has been modified}. Not changing the required fields and signature in the certificate would result in RPs discarding the object before processing its content due to failing signature validation. These objects thus need to be internally adapted by CURE instead of simply adding a EE-certificate as in the case of content objects.\\ 
\indent Adapting fields inside objects that are generated to fuzz the RPs is non-trivial. If CURE changes values inside the object, it might overwrite fields purposefully changed to test the reaction of the RPs. For example, the generation algorithm might intentionally change values in the signature to probe the reaction of the RPs to an invalid signature value. Overwriting the signature would hinder the test case. However, if the generated object manipulated other fields but not the signature, the RPs will discard the objects if CURE does not correctly sign the object. The same logic applies to a number of fields in the certificate and CRL. In total, 8 fields are interdependent on the repository configuration and need to be valid to ensure the RPs parse all certificate fields. These fields include the signature, parent key identifier, CRL location, issuer name, issuer location, repository location, manifest location, and the notification location. However, replacing any of these fields might override an intentional change.\\ 
\indent {\bf Replacement strategies:} To solve this problem, CURE allows for different replacement strategies, including optimistic replacement, targeted replacement, parseable replacement, and no replacement.\\
\indent $\bullet$ Optimistic replacement replaces all fields that need to be adapted. This configuration is generally not recommended as it only allows to test fields which are not replaced, limiting the capability of the tool.\\
\indent $\bullet$ Targeted replacement replaces all fields that are flagged to be replaced by object generation. This identification is built into the generation algorithm. If a field should be replaced, the generation algorithm inserts the field with a none value, allowing CURE to only replace fields which are not manipulated. Targeted replacement is the default mode for structure aware object generation. This approach is not possible in random mutations of the entire object, as in this case, mutations do not target a specific field. Further, replacing fields is only possible if the object can be parsed by CURE, which in many cases is not possible with randomly mutated objects.\\
\indent $\bullet$ Parseable replacement changes all fields that are parseable by CURE, i.e., fields that are well formatted. The idea of this strategy is that mutated fields are usually not well formatted. Thus, if the field does not look correct, CURE concludes that the field has likely been manipulated by the object generation algorithm and leaves the field intact. This is the default replacement strategy for randomly mutated objects.\\
\indent $\bullet$ No replacement leaves all fields intact. This circumvents the risk of overwriting a field that was manipulated. However, the RPs will not process the object completely because the signature of the object does not fit the CA of CURE.

\subsubsection{Scaffolding the Repository} After the object has been adapted and signed, it has to be inserted into the repository with the required auxiliary objects to feed it to the RPs. This process depends on the object type. Since we aim not only to find crashes but also to identify inconsistencies, CURE always creates a complete repository, including ROAs. By adding ROAs, CURE can identify if a certain object was accepted by the RPs. For example, consider a certificate where certain fields lead to some RPs discarding the object. If the certificate does not link to any ROAs, CURE cannot easily recognize which RP accepted the certificate. However, if ROAs are added, CURE can compare the accepted ROAs of each RP and find the inconsistency.\\
\indent An additional challenge in adding the objects to a repository are restrictions in the amount of certain objects per CA. While it is possible to place all ROAs, GBRs or ASPAs in a single CA, the RPKI demands only a single Manifest and CRL file per CA. Thus, to test multiple Manifests or CRLs, CURE needs to create a corresponding amount of new CAs, each including the appropriate objects, i.e., a Manifest, CRL and a ROA file to check for inconsistencies. The same logic applies to CA certificates, as each certificate needs to map to a single repository folder. For RRDP files, the effort increases further as the RFC dictates only a single Notification and Snapshot file per PP. Thus CURE emulates a setup with a corresponding number of CAs, with each certificate pointing to its own generated Notification file within a unique domain. The same setup applies for Snapshots.\\ 
\indent Since scaffolding requires a lot of computational effort, CURE caches the majority of the required auxiliary objects and adapts them to the inserted generated objects. 

\subsection{Interacting with the RPs}
After creating an RPKI repository around manipulated objects, CURE executes the RPs to stress test their validation. For direct control over the execution time, the RPs are run in single-validation mode, i.e., they are executed and terminate after finishing one round of validation. This setup necessitates a crash detection algorithm that does not rely on a simple monitor that probes if the process is still active. To probe for crashes, CURE uses two indicators. First, the exit code of the RP process is checked after the RP process terminates. A non-zero exit code indicates that the process did not finish as expected, indicating a potential crash during execution.\\
\indent Second, the distributor checks the output files of the RPs. Because the RPs write the final output to disc only after validation, a crash during execution will result in no output of the RP. If no VRP file was created after the process exits, a potential crash is logged. During the course of our experiments, we did not find any false positives with our crash detection logic. Once a crash is detected, CURE runs additional logic to identify which exact object in the batch caused the crash. It then generates a detailed crash report, describing object content, RP logs and time of the crash. A similar process is also applied to inconsistent validation results. If all RPs produce VRP files, i.e., do not crash, CURE parses the files and checks for inconsistencies in the accepted ROAs. If inconsistencies are identified, a text report detailing the nature of the inconsistency, affected RPs and trigger object is generated.

\begin{table*}[t!]
\scriptsize
\renewcommand{\arraystretch}{0.7}
    \centering
  \begin{tabular}{l|c|c|c|c|c|c|c|c|c|c|c|c|c}
    \multirow{2}{*}{\textbf{Taxonomy}} &
      \multicolumn{4}{c|}{\textbf{Coding Errors}}\label{cod_errors} &
      \multicolumn{1}{c|}{\textbf{Cache Disparity}} &
      \multicolumn{7}{c|}{\textbf{RFC Non-Compliance}}\\
      \cline{2-13}
    & \scriptsize{\makecell{Path \\ Traversal}} & \scriptsize{\makecell{{RTR} \\ {Buffer} \\ {Overflow}}} & \scriptsize{\makecell{{Decoding} \\ {ROA} \\ {IP Resources}}} & \scriptsize{\makecell{{Decoding} \\ {Malformed} \\ {ASN.1}}}
    &  \scriptsize{\makecell{PubPoint \\ DoS}} & \scriptsize{\makecell{[6482] \\ \S 3}} & \scriptsize{\makecell{[6487] \\ \S 5}} & \scriptsize{\makecell{[9286] \\ \S 5}} & \scriptsize{\makecell{[6487] \\ \S 4}} &  \scriptsize{\makecell{[8182] \\ \S 3}} & \scriptsize{\makecell{[6482] \\ \S 4}} & \scriptsize{\makecell{[8897] \\ \S 4}} \\\hline
    Routinator & \cmark & \xmark & \cmark & \cmark & \cmark &  \cmark  & \xmark & \cmark & \cmark & \xmark& \xmark & \xmark\\ 
    Fort & \xmark & \cmark  &  \xmark & \xmark & \cmark &   \xmark& \xmark & \cmark & \xmark & \xmark& \xmark& \cmark\\
    OctoRPKI & \xmark & \xmark  &\xmark  & \cmark & \cmark &   \cmark& \cmark &  \cmark & \cmark & \cmark& \cmark& \xmark\\
    rpki-client & \xmark &  \xmark  & \xmark & \xmark & \cmark &   \cmark & \cmark & \xmark & \xmark & \xmark& \xmark & \xmark\\    
    \hline
    
    \textbf{Artifacts} & & & & & & & & & & & & &\\
    ROA & \xmark &- &\cmark &\cmark&\cmark&\cmark&\xmark& \xmark & \xmark & \xmark & \cmark & \cmark\\
        MFT & \xmark &- &\xmark &\cmark&\cmark&\xmark&\xmark& \cmark & \xmark & \xmark &\xmark&\cmark\\
        CER & \xmark &- &\xmark &\cmark&\cmark&\xmark&\xmark& \xmark & \cmark & \xmark &\xmark&\cmark\\
        CRL & \xmark &- &\xmark &\cmark&\cmark&\xmark&\cmark& \xmark & \cmark& \xmark &\xmark&\cmark\\
        Snapshot & \xmark &- &-&-&\cmark&-&-& -&-& \cmark &\xmark& \xmark\\
        TAL & \cmark &- &-&-&-&-&-& -&-& \xmark &\xmark& \xmark\\
        RTR PDU& \xmark &\cmark&-&-&-&-&-& -&-& \xmark &\xmark&\xmark
  \end{tabular}
  \vspace{-5pt}
  \caption{Errors, inconsistencies and crashes we found in popular RP implementations.}
  \vspace{-15pt}
  \label{tab:1}
\end{table*}
\subsection{Improving Performance of the CURE}
\begin{figure}[t!]
    \centering
    \begin{subfigure}{0.47\columnwidth}
        \includegraphics[width=\linewidth]{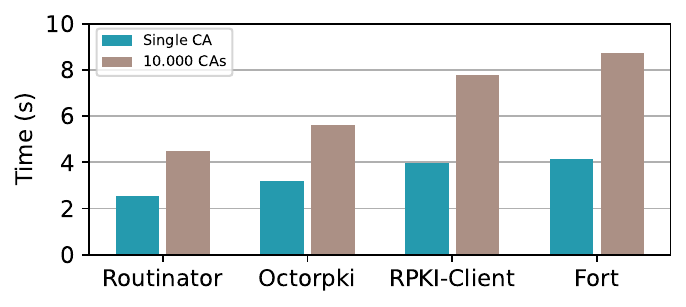}
    \end{subfigure}
    \hspace{-5pt}
    \begin{subfigure}{0.47\columnwidth}
        \includegraphics[width=\linewidth]{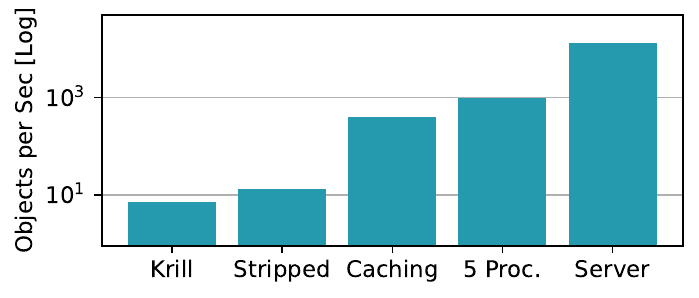}
    \end{subfigure}
     \vspace{-5pt}
    \caption{Execution time for 10000 objects of investigated RPs (left) and processing capability of CURE in different optimization stages (right).}
    \label{fig:comparison}
    \vspace{-15pt}
\end{figure}

The execution time of CURE is of major concern for fuzzing, as fast execution allows us to test more objects per timeframe. To speed up the tool, we first identify speed limitations in the publication pipeline of RPKI objects. We then optimize the structure of CURE to significantly improve its processing speed.\\
\indent To understand the performance slowdown, we first measure the execution speed of the most popular implementation of an RPKI publication point, Krill. Krill allows for the bulk creation of ROAs and is thus well suited to provide a baseline execution time from which we will optimize CURE. To test Krill's execution time, we use the \textit{krillc roas update --delta} command for the bulk creation of 1000 ROAs. The execution takes 122.7s, which corresponds to 122.7ms per ROA. Using the ROA creation of Krill as the underlying implementation for CURE would thus allow the fuzzer to test a maximum of 8.1 ROAs per second.\\
\indent {\bf Unnecessary operations.} We test the execution time of our basic ROA creation implementation. In our implementation, we removed the majority of checks that Krill uses to ensure no faulty ROAs are added to the repository. These checks limit both the speed and flexibility of the fuzzer in processing malicious objects. With the removal of the checks, CURE reaches a speed of 77.0 seconds for 1000 ROAs, i.e., 77ms per ROA, allowing for 12.98 ROAs per second. This is still insufficient for the operation of the fuzzer. 
To find slow sections of the program, we use Linux perf.\\
\indent {\bf Keys generation time.} Analysis with perf identifies that the majority of processing time is spent on the creation of RSA keys for the created ROAs. In the Rust implementation of OpenSSL, used both by Krill and CURE, we measured the average creation time of a single 2048-bit RSA key pair to be about 70ms. In the normal operation of the RPKI, this overhead is necessary, as each key needs to be unique to ensure security, which is not relevant for CURE. It is thus possible to reduce this overhead by pre-calculating and caching a batch of unique RSA keys. In each batch execution of CURE, the keys are loaded from the cache, removing the overhead of key creation. With the caching of the keys, the execution is sped up significantly; the creation of 1000 ROAs takes 2.25s, 2.25ms per ROA. Thus operating the fuzzer with cached keys allows for testing of 400 objects per second.\\
\indent To find additional potential for optimization, we use perf again to identify slow functions in the code. However, analysis shows that 85\% of execution time is spent on signature generation. Since the RPs require a valid signature on each object and the signature depends on the content of the object, caching signatures is not possible and hence speed up of object generation is not possible.\\
\indent {\bf Parallel processing.} The second approach to improve CURE's execution time is parallel processing. Since the creation and signing of each object is an atomic operation, parallelization can provide additional speed-up. This parallelization is achieved by spawning multiple processes that handle the most computationally intensive tasks: parsing and signing objects. The processes are spawned to run autonomously, reading batches of objects, parsing and signing them, creating scaffolding RPKI objects and serializing all created objects. CURE continuously reads the serialized file, writes the objects to the local repository and creates the required RRDP files to distribute the repository content to RPs. To test the speed-up by multiprocessing, CURE is executed with 5 signing processes. This results in a processing time of $<$1ms per object, translating to over 1000 objects per second.\\
\indent {\bf Hardware optimization.} The final optimization is the hardware used to execute CURE. Since the design of CURE utilizes multiprocessing, more computing cores allow for further speed-ups of the execution. Improved single-core performance also decreases the time required for object processing. For the speed evaluation, CURE is executed with 20 processes on a computing server, equipped with two XEON Platinum 8380 processors with a clock frequency of 2.3 GHz and 40 physical cores. This results in a speed of 13000 objects per second. An overview of the different  (logarithmic) speed values is plotted in Figure \ref{fig:comparison}.
To evaluate if this speed is sufficient for the fuzzer, we also investigate the processing speed of the RPs. If CURE is able to reach a processing speed that surpasses the validation speed of the RPs, we eliminate idle RP time. The processing time of the RPs is tested with a setup of 10.000 signed ROAs. Two cases are tested: all ROAs signed by a single CA and every ROA signed by a separate CA. The results are shown in Figure \ref{fig:comparison}. The fastest execution time is measured to 2.2 seconds for the single CA case in Routinator. In its fastest configuration Routinator can thus process 4545 ROAs per second. This is significantly slower than the speed at which CURE can create valid repositories.\\
\indent In a multi-RP setup, the maximum speed of testing objects is dictated by the slowest RP, which according to our measurements is Fort. Since Fort cannot process more than 2500 objects per second, the maximum speed of the fuzzer cannot surpass this amount, which corresponds to roughly nine million objects per hour.
\subsection{Limitations}
CURE is not guided, i.e., it does not get feedback from the RP binaries informing it which execution paths were triggered by a specific object. This limitation directly results from the language agnostic design; if a fuzzer should be usable with every RP, it cannot rely on manual compilation of the RP binaries to insert the required flags for coverage guided fuzzing. The design of CURE is thus a tradeoff between usability and execution efficiency. While conventional fuzzers are able to use guided techniques to optimize for execution paths in the investigated binary, their setup is very labor-intensive and language specific. CURE lacks coverage detection capabilities, but makes up for the limitation with a high execution speed and low operational effort to setup.

\section{\hspace{1mm}Taxonomy of Errors in RPKI}\label{taxonomy}
Our experiments with CURE and subsequent code analysis of the RP implementations reveal a variety of issues in the internal logic, ranging from coding errors to inconsistency with RFC standards, to subjective interpretation of vague RFC directives. In total, we found 19 critical vulnerabilities in the RPs, all leading to a crash of the RP, with one issue allowing poisoning of the trust anchor cache. Further, we discovered numerous problems leading to RPs reaching different validation results on the same RPKI repositories, thereby ultimately supplying routers with inconsistent VRPs. The lack of uniformity in the internal data processing of RPs can disable RPKI protection for the prefixes covered by the corresponding ROAs, therefore allowing BGP hijacks to take place despite victims having issued ROAs via valid channels. Table \ref{tab:1} provides an overview of the discovered vulnerabilities in the four most popular RP implementations on the Internet. In the following, we describe the core characteristics of the vulnerabilities we discovered, how they can be exploited and the consequences of their exploitation for global routing security.

\subsection{Fuzzing Setup}
To find problems in the RP implementations, we used CURE to test RPs on all standardized RPKI objects. For each object, we ran CURE for a total of 8 hours resulting in approx. 72 million test cases. We deemed this time as sufficient as we never discovered vulnerabilities after more than 4 hours on any object. This resulted in a total fuzzing time of 72h for 648 million total test cases.

\subsection{Coding Errors in RPs}
Our fuzzing campaign yielded a multitude of crashes. The cause of all severe vulnerabilities that we found are coding errors in the RP implementations. The majority of these errors are caused by the lack of proper error handling for malformed object fields. Further, we also find errors caused by a lack of proper sanitization of user controlled input to store downloaded objects, leading to a path traversal attack.

\subsubsection{Path Traversal: RPKI Poisoning}
\begin{figure}[t!]
    \centering
    \includegraphics[width=0.95\columnwidth]{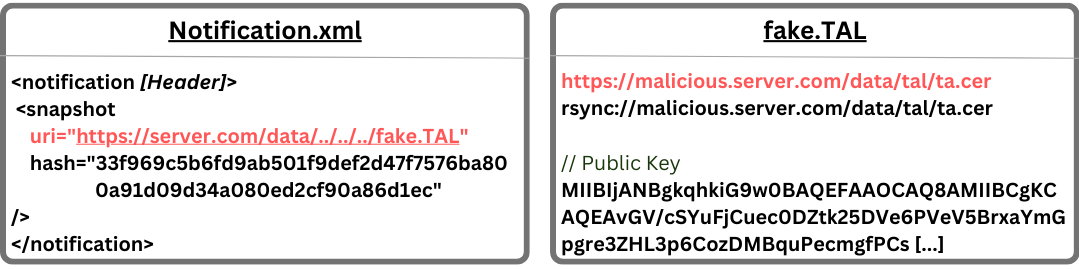}
    \vspace{-5pt}
    \caption{RPKI Poisoning Attack via path traversal.}
    \label{fig:notification_snapshot}
    \vspace{-20pt}
\end{figure}

Our analysis discovered a path traversal attack in the object storage module of Routinator, caused by a lack of filtering for malicious characters while processing URIs. The attack can only be triggered when the parameter \textit{RRDP-keep-responses} is enabled, which directs Routinator to store downloaded RRDP files to a directory. The exact storage location is deduced from the URI of the RRDP object, controlled by the PP itself. The URI parsing which determines the exact storage location for RRDP objects does not conduct checks for path traversals, making it vulnerable to a path traversal attack. The vulnerability is only present in the URI parsing of HTTPS addresses. The method that does rsync URI parsing specifically checks for path traversal attacks and escapes the input. 
No such check is present for HTTPS URIs. For an interested reader, the referenced code parts are in Appendix, Listings \ref{fig:rrdp-pt}, \ref{fig:rsync-pt}, and \ref{fig:rsync-es}. Since rsync is only used by the RPs during failures with RRDP as a backup, a vulnerability in RRDP is more severe and immediately affects all Routinator instances.\\
\indent {\bf Exploit.} We evaluated the path traversal vulnerability with an adversary that controls a PP in the RPKI tree. Controlling a malicious PP is easy, e.g., if the adversary owns an AS it can add a delegation to its PP from the corresponding RIR. To exploit the vulnerability, the adversary needs to add a malicious payload in their \textit{notification.xml} file. As Routinator utilizes object URIs as file paths and stores the objects wherever the path specifies, the attacker can deliver malware on Routinator's host or sabotage RPKI itself. An example for RPKI sabotage is shown in Figure \ref{fig:notification_snapshot}. An attacker creates a \textit{fake.TAL} which points to non-RIR certified publication points. Via the path traversal, the attacker delivers the malicious payload to the TAL folder of the Routinator installation during an RRDP fetch. Routinator then loads the attackers' TAL in the next validation run, follows the inherently trusted malicious certificate and validates malicious ROAs. The attacker can thus get arbitrary prefix-AS pair validated by Routinator and thereby propagate fake ROAs. Since Routinator does not raise any alarms when the TAL folder changes, this attack is difficult to detect, allowing a malicious entity to circumvent RPKI protection without detection. This attack is possible for Routinator versions before 0.12.0. From this version onwards, Routinator changed the TAL management, which mitigates the attack scenario. However, the underlying vulnerability of path traversal is still present even in newer releases, still allowing the delivery of malware to the system.

\subsubsection{Crashes: RPKI Downgrade Attacks}\label{sc:crashes}
Our analysis of RPs discovered several DoS vulnerabilities for 3 out of 4 main RP implementations: Routinator, Fort, and OctoRPKI. These DoS vulnerabilities can be triggered in different sections of the code and cause a total crash. 
Since the RPs are required to contact all the RPKI PPs during a refresh interval, an attacker with a valid PP can potentially crash all vulnerable RP instances from a single point of attack. In Table \ref{tab:1} Section ``Coding Errors'', we list an overview of the discovered crashes and the affected RPs.
We observe that Routinator and OctoRPKI can be DoS-ed by any external PP, whereas Fort is vulnerable to memory allocation issues given misconfigured packet sizes. We explain the bugs with code snippets in Appendix \ref{bugs}.\\
\indent \textbf{Routinator} is a Rust implementation of an RP. It relies on many same-vendor created libraries to perform most necessary parsing, validation, and computation. Our analysis has discovered that Routinator is prone to crashes that are triggered in the RP source code as well as in the external same vendor libraries that support Routinator and Krill functionalities. Specifically, our analysis discovered that libraries, such as \textit{rpki-rs} and \textit{bcder}, show major problems in their implementations due to the use of assertions, lack of sanity and length checks, lack of comprehensive error handling and deliberate disregard to handle corner cases. These libraries are used by both Krill and Routinator. 
We discovered in total 11 bugs for Routinator, all of them triggered in different parts of the code by different types of mutated RPKI objects. Table \ref{tab:1} shows an overview of all discovered crashes, 2 of which can be found in \textit{rpki-rs}, 8 in \textit{bcder} and 2 in the \textit{routinator app} source code. \\
\indent \textbf{OctoRPKI} is implemented in Go. All its RPKI related functionalities are self contained, so the bugs can be found in the source code of the application itself. OctoRPKI, same as Routinator, has implemented its own parsing and validation libraries. The code is missing sanity and lengths checks, leading to crashes when malformed files are fed to the system.\\
\indent {\bf Fort}, unlike Routinator and OctoRPKI, is much harder to compromise. 
\begin{figure}[t!]
  \centering
  \begin{minipage}[t]{0.22\textwidth}
    \centering
    \includegraphics[width=\linewidth]{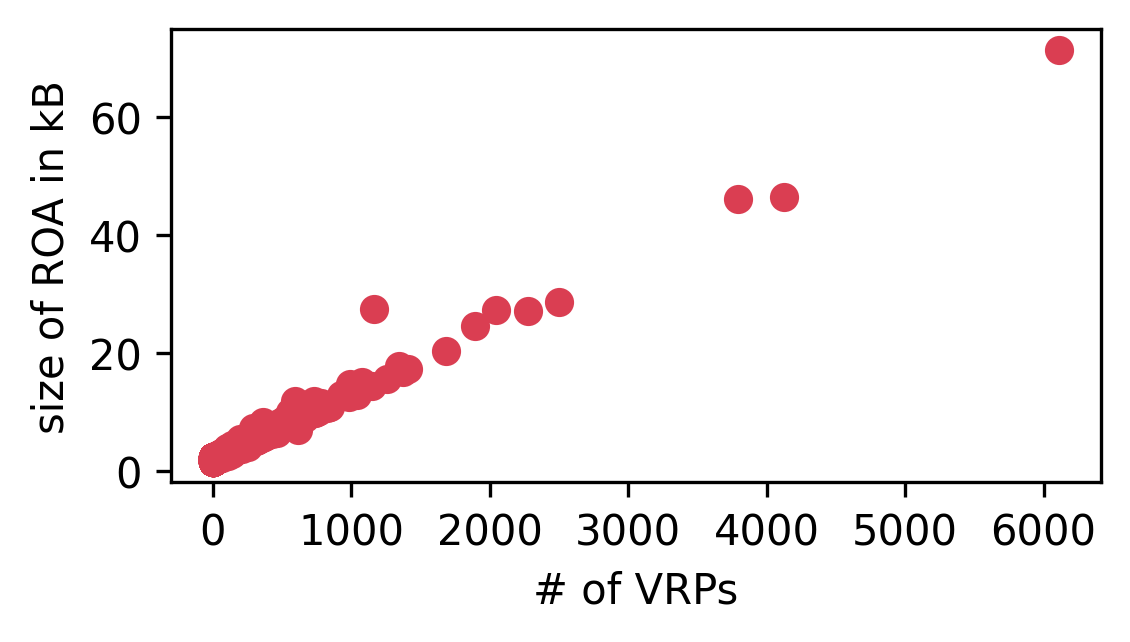}
    \vspace{-15pt}
  \end{minipage}
  \hfill
  \begin{minipage}[t]{0.255\textwidth}
    \includegraphics[width=\linewidth]{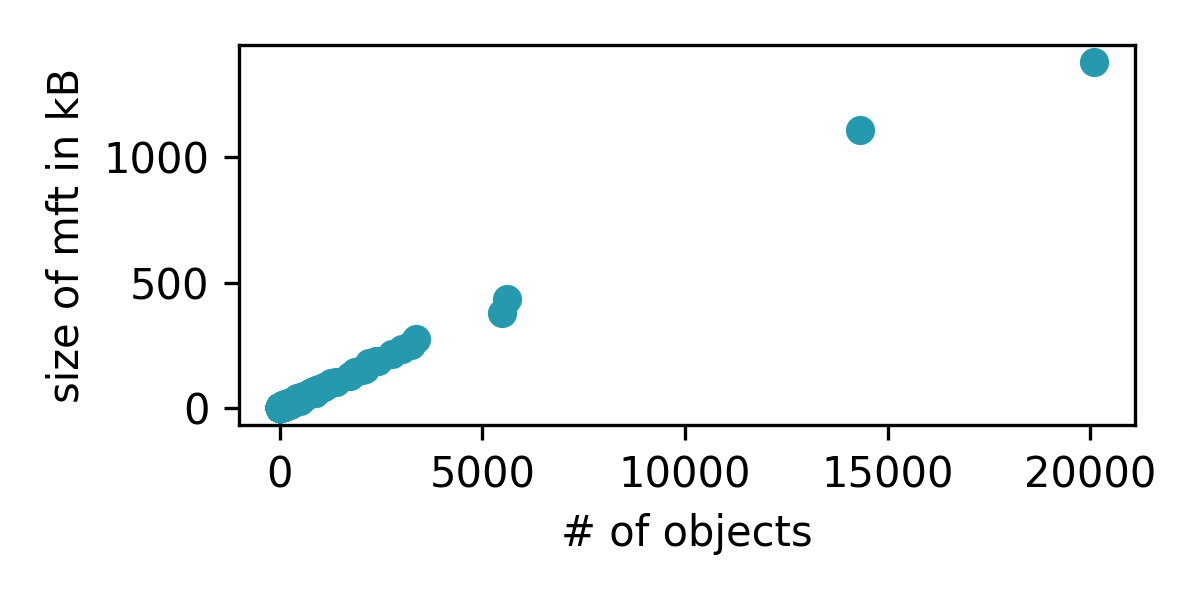}
    \vspace{-15pt}
  \end{minipage}
  \caption{ROA and Manifest sizes and payload.} 
\label{fig:manifest}\label{fig:roas}
\vspace{-5pt}
\end{figure}
However, we found a DoS vulnerability in their RTR server, i.e. the server that interacts with connected BGP routers to exchange validated RPKI information. An attacker addressing the RTR server of Fort can crash the RP application by sending a message larger than the expected Protocol Data Unit (PDU) slot. The result is a memory corruption that leads to a full application crash, disabling RPKI validation.\\
\indent {\bf Exploitation.} We find experimentally that as a result of the crashes caused by the DoS vulnerabilities, the RPs are no longer in a live session with the BGP border routers, eventually leading to the routers flushing their RPKI cache and not applying RPKI validation to newly received BGP updates. The holdout period that the cache remains valid after the RTR connection is dropped varies on a per router basis. We measured the default flushing periods for Cisco \cite{cisco:rpki} to 360 seconds, Juniper \cite{juniper:rpki} to 3600 seconds and FRR \cite{frr:rpki} to 7200 seconds. Thus, if the RPs are not manually restarted withing two hours of the DoS attack, the routers will not apply RPKI validation to inbound BGP updates anymore, effectively leading to a downgrade of RPKI validation. The DoS attacks can be triggered by a malicious PP anywhere in the RPKI tree. The size or location of the malicious PP is irrelevant as long as the PP is reachable by the victim. As any entity owning IP resources is eligible to host their own PP, this attack against RPs is realistically exploitable even by a small-scale adversary.

\begin{table}[b!]
\scriptsize
\renewcommand{\arraystretch}{0.7}
\centering
\vspace{-5pt}
\begin{tabular}{l|c|c|c|c|c}
RP & ROA / MFT & CRL & CERT & ASPA & GBR  \\ \hline
Routinator & 20MB & 100MB & 5MB & 20MB & 48MB\\ 
Octorpki & 1.9GB  & 700MB & 5MB & 1.9GB & 1.9GB\\
Fort &  7MB & 10MB & 5MB & 10MB & 10MB\\
rpki-client & 4MB & 4MB & 5MB & 5MB & 5MB\\
\end{tabular}
\vspace{-5pt}
\caption{Single file size to crash snapshot.xml parsing.}
\label{tab:3}
\vspace{-5pt}
\end{table}

\subsection{Cache Disparity}
We observe that feeding RPs with validly signed large objects can cause the snapshot parsing to fail. The parsing failure is independent of the snapshot.xml size and is triggered if single files are above some arbitrary threshold. The thresholds in question can vary from 4 MBs for rpki-client, to 1.9 GB for OctoRPKI, see Table \ref{tab:3}. 
This error results in a ``silent'' discarding of individual PP repositories. The RP itself does not crash and continues to periodically fetch the RPKI data, but any object inside the malfunctioning PP will be discarded by the RP. This issue can cause discarding of valid objects and thereby downgrade of protection both organically and through an attacker. The attack angle can be leveraged against repositories hosting multiple entities, such as RIR repositories. An attacker can then inflate the count and size of its objects, for example by adding a massive number of entries to a ROA or a manifest, leading to big file sizes in the snapshot of the hosting PP and therefore cause the discarding of the entire repository by some or all RPs. This problem can also arise accidentally when an institution creates large RPKI objects to cover their resources. Further, with the growing number of utilized IPv6 IP address space, we expect that the number of RPKI resources will grow, and correspondingly both ROAs and manifest files will increase in size in the future.\\
\indent {\bf Exploit.} The attack exploiting this vulnerability is similar to the one in Section \ref{sc:crashes}, essentially disabling RPKI validation for the affected prefixes. For large repositories hosting many different entities, this attack can have a detrimental effect on routing security, as the objects of all other parties in the same PP will not be used to validate BGP updates anymore.\\
\ignore{
\begin{figure}[h!]
\centering
    \includegraphics[width=0.35\textwidth]{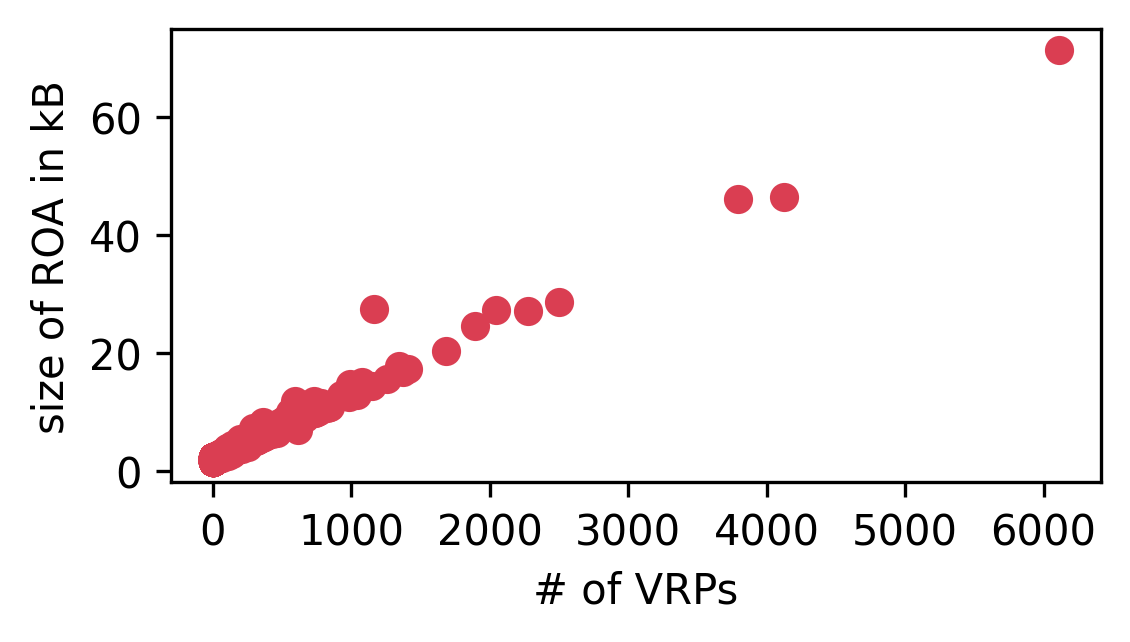} 
    \includegraphics[width=0.35\textwidth]{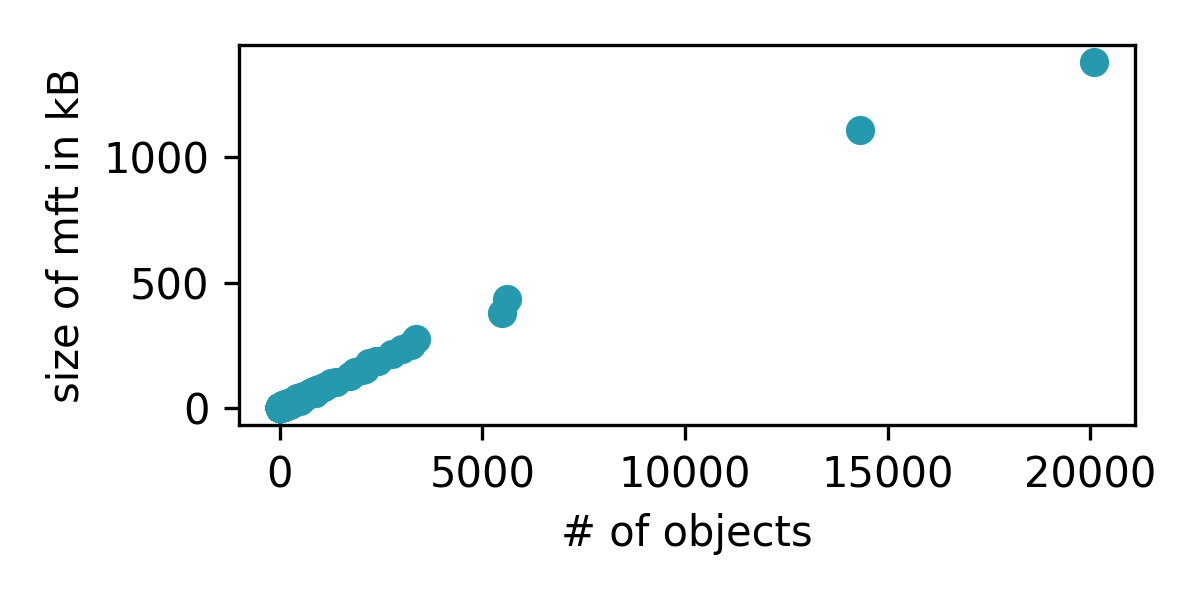} 
    \caption{ROA and Manifest sizes vs. payload.}
    \label{fig:payload}
    \vspace{-15pt}
\end{figure}
}
\indent {\bf Impact now and in the future.} We measure the payload and file sizes in kilobytes for ROAs and manifests as shown in Figure \ref{fig:roas}. 
The largest files in the RPKI cache are manifests, with the largest manifest being 1.4 MB and listing 19,958 artifacts. The average boilerplate size of MFTs is 2078 bytes, and entries have an average size of 70 bytes each. For a manifest to reach at least 4 MB in size, the lowest file size that would cause problems processing MFTs (see Table \ref{tab:3}), it would need to contain on average 59889 objects of 70 bytes each. 
In Figure \ref{fig:mftsize} we see an evaluation of the number of entries per manifest necessary to discard the snapshot parsing of each RP: rpki-client has the lowest threshold as it only needs 59,889 objects, next is Fort with 104,828 objects followed by Routinator with 299,564. OctoRPKI is by far the most resilient implementation needing an average of 29 million entries to reach the discard size of 1.9 GB. Thus the most vulnerable RPs, in order of low threshold, are: rpki-client, Fort and Routinator. Table \ref{tab:3} provides a detailed summary of all the different thresholds for each RP implementation and file type that can trigger this issue. Given the projected growth of RPKI repositories as well as the addition and adoption of new record types (ASPA, BGPSec Certificates etc.) we project that the size of a manifest will organically grow 3-4x the existing size in the short term to accommodate more objects. Manifests have to list all RPKI objects contained in a valid CA repository, therefore given the variety of objects and the fact that we are only at 44\% global RPKI deployment \cite{nist}, it is expected that the repository size of large providers will increase considerably. This error is particularly substantial for RPKI providers, such as the RIRs, which are at the forefront of expanding their databases with new objects and new LIRs. It is thus vital that repositories adapt size thresholds in the future to accommodate for the natural growth of the RPKI ecosystem. 

\ignore{
\begin{figure}[h!]
\centering
    \includegraphics[width=0.8\columnwidth]{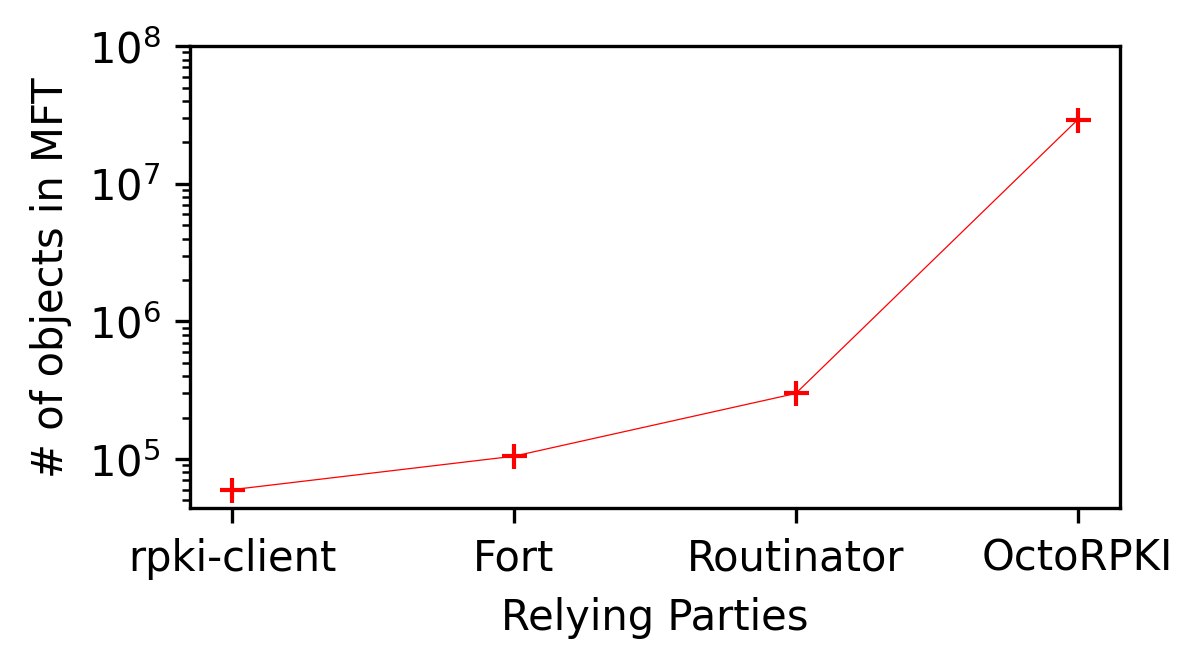} 
    \caption{\# of MFT entries to crash Relying Parties.}
    \label{fig:mftsize}
    \vspace{-10pt}
\end{figure}

}

\begin{figure}[t!]
  \centering
  \begin{minipage}[t]{0.23\textwidth}
    \centering
    \includegraphics[width=\linewidth]{figures/mft-payloads.png}
    \vspace{-15pt}
\caption{\#MFT to crash RPs.} 
\label{fig:mftsize}
  \end{minipage}
  \hfill
  \begin{minipage}[t]{0.23\textwidth}
    \includegraphics[width=\linewidth]{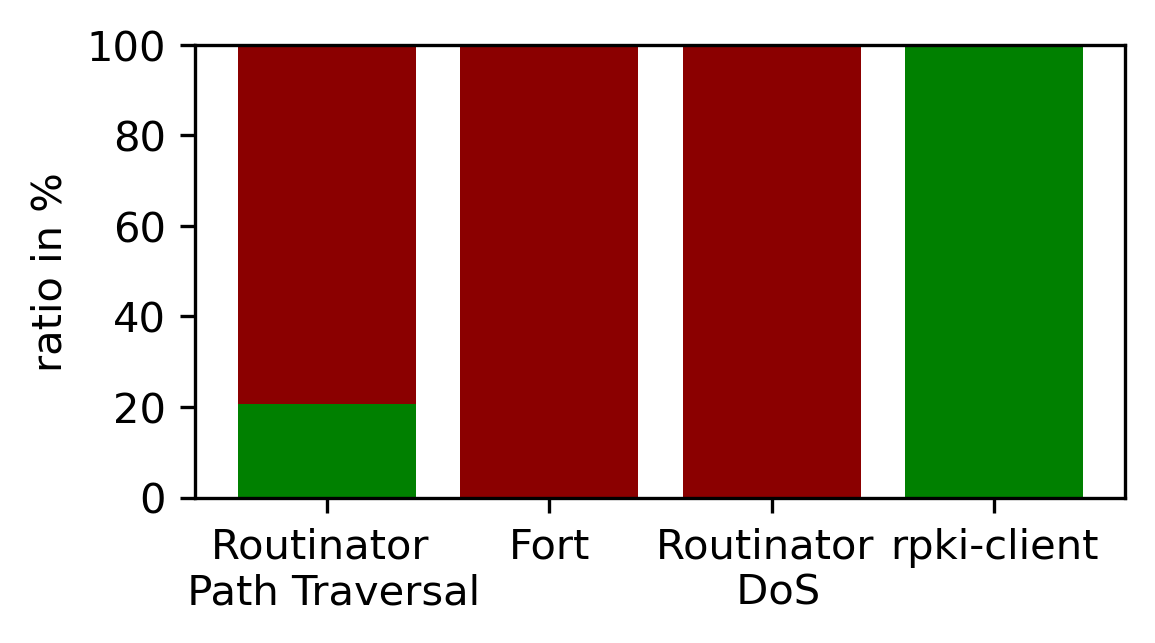}
    \vspace{-15pt}
\caption{Vulnerable RPs.}
    \label{fig:ratio}
  \end{minipage}
\vspace{-25pt}
\end{figure}

\subsection{RFC Inconsistencies}
Our analysis finds that, in many instances, RPs are also RFC non-conforming. The non-conformity manifests either as strictly defined validation and parsing checks which are not respected, leading to security issues, or as undefined corner cases which are handled differently by the RPs. We find that these RP idiosyncratic behaviors lead to discarding of valid ROAs. The lack of a systemic deterministic behavior affects ROA propagation worldwide and leads to a silent downgrade of RPKI protection for the affected regions of the Internet, despite the corresponding prefixes having issued ROAs.\\
\indent {\bf Objects processing.} Manifest is one of the core files that define the data parsing and processing for RPs. The standard for manifest generation and processing defined in [RFC9286] specifies in Section 6.6 Appendix B a mandatory presence of a valid CRL for every manifest. This update was issued in 2019 to address security issues in the previous, now obsolete [RFC6482]. According to our observations, OctoRPKI has not updated its processing to the recent standard and still serves the content of a manifest with an expired CRL. Since CRLs are so important for a manifest's processing, we investigated the robustness and uniformity of CRL processing. According to [RFC6487] there are several requirements for the CRL format and processing; e.g., CRLs must contain two mandatory extensions (CRL Number and Authority Key Identifier). OctoRPKI and rpki-client ignore a missing extension. Furthermore, we observe that  missing non-optional CRL fields \textit{signature}, \textit{signatureAlgorithm} or \textit{issuer} do not stop the CRL from being processed, and the manifest is accepted for OctoRPKI. Additionally, Routinator is the only implementation which requires a much stricter formatting of the CRL number extension than all other parties. We further evaluated the parsing of snapshots and ROAs. OctoRPKI is the only RP that does not check concurrency of the \textit{session\_id} parameter in the notification and snapshot files, thus parsing files with inconsistent IDs against the requirements of [RFC8182]. This error exposes the affected RPs to replay attacks.\\
\indent {\bf ROA parsing.} Parsing of ROAs is a vaguely specified part of the RFC. The ROA-specific standard [RFC6482] and its successor [RFC6488] have few specifics about how to treat ROAs and corner cases. We discovered two main inconsistencies resulting from undefined cases in the RFC. When non-critical ROA schema fields such as \textit{MaxLength} contain non-integer content, all RPs except for OctoRPKI will drop that ROA. Additionally, the format of \textit{ipAddrBlocks}, the ROA field where the prefix-AS pairs are stored, can be provided two ways as depicted in Figure \ref{fig:roa} in Appendix. Routinator has a very strict parsing rule that no other RP supports. According to the strict parsing, the \textit{ipAddrBlocks} list can only contain two entries, for IPv4 or IPv6. The general parsing method as implemented by rpki-client, OctoRPKI and Fort allows for multiple entries with the same IP family within the field. [RFC6482] references the format to [RFC3779], which explicitly defines the field as a sequence, not a set. This implicitly allows for multiple fields of the same IP family. For a more detailed enumeration of the RFC inconsistencies refer to Appendix \ref{rfc-inconsistency}.

\begin{table}[h!]
\scriptsize
\renewcommand{\arraystretch}{0.8}
\centering
\vspace{-5pt}
\begin{tabular}{l|c|c|c}
{\bf Routinator} & {\bf OctoRPKI} & {\bf Fort} & {\bf rpki-client}  \\ \hline
0.13.0-dev & v1.5.10 & 1.5.3 & 8.2 \\
\end{tabular}
\vspace{-5pt}
\caption{Tested relying party versions.}
\label{rp-versions}
\vspace{-15pt}
\end{table} 
\section{\hspace{3mm} Extent of Vulnerabilities on the Internet}\label{sc:impact}

We evaluate the extent to which networks on the Internet are affected by the vulnerabilities found in Section \ref{taxonomy}. The evaluations in Section \ref{taxonomy} were conducted using the latest RP versions, listed in Table \ref{rp-versions}. For the analysis, we log all RP instances querying publication points in April 2023 and extract their version number from their agent header.

\subsection{Impact of Attacks}
\begin{figure}[t!]
  \centering
  \begin{subfigure}[t]{0.17\textwidth}
    \centering
    \includegraphics[width=\linewidth]{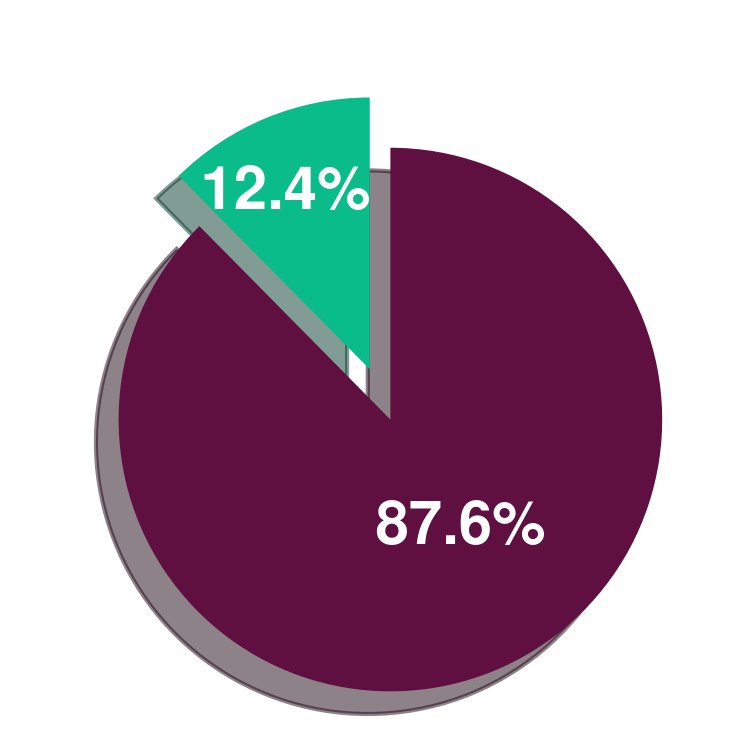}
    \vspace{-20pt}
    \caption{\footnotesize{Vulnerable ASes.}}
    \label{fig:asn-vul1}
  \end{subfigure}
  \hspace{4em}
\begin{subfigure}[t]{0.17\textwidth}
    \includegraphics[width=\linewidth]{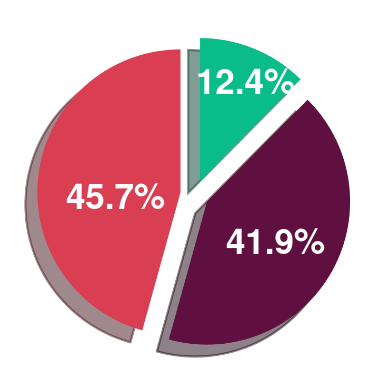}
    \vspace{-20pt}
    \caption{{\footnotesize{Path-Traversal \& DoS.}}}
    \label{fig:asn-vul2}
\end{subfigure}
\vspace{-10pt}
\caption{Ratio of ASes vulnerable to RPKI downgrade.}
\label{fig:asns-affected}
\vspace{-15pt}
\end{figure}

{\bf Vulnerable RP deployments.} In our evaluation, we found 18 critical vulnerabilities in the RP implementations.
Figure \ref{fig:ratio} shows a breakdown of the ratio of RP instances deployed on the Internet that are vulnerable to our attack vectors. The RPKI poisoning attack on Routinator affects 79.4\% of all Routinator instances. The remaining 20.6\% instances are vulnerable to the path traversal but not to the poisoning attack. 
Malicious PPs can still store custom payloads on the host's drive. 
All Routinator, OctoRPKI and Fort deployments are vulnerable to the discovered DoS attacks (Table \ref{tab:1}). So far, we haven't found any vulnerabilities on the rpki-client.\\
\indent {\bf Affected ASes.} 
We approximate the ratio of RPKI-protected networks that would lose RPKI protection in case an adversary exploits the crashes in Table \ref{tab:1}. Figure \ref{fig:asns-affected} describes the ratio of ASes immediately vulnerable to our attacks. According to Figure \ref{fig:asn-vul1} only 12.4\% of all ASes are safe from path traversal and crashes of RPs. The remaining 87.6\% of ASes can have their RPKI protection downgraded. In Figure \ref{fig:asn-vul2} we show that 45.7\% of the vulnerable ASes are affected both by RPKI poisoning and DoS attacks. The remaining 41.9\% are vulnerable only to DoS attacks. Overall, RPKI protection of at least 354,869 prefixes can be disabled.\\
\indent We also found validation inconsistencies, that led to validation discrepancies in different RPs. We found these inconsistencies by comparing VRP entries generated by the different RPs. We further ran all RPs against the real-world RPKI data to find inconsistencies already manifesting on the Internet. We confirmed the existence of VRP inconsistencies if the VRP sets of all RPs are not identical. 

\subsection{Vulnerable Prefixes During Benign Network Conditions}
{\bf Discrepancies in VRPs.} Previous work \cite{friedemann2022assessing} attributed differences in real-world VRPs' entries to errors in the communication between the PPs and the RPs, but not to processing inconsistencies. However, our investigation shows that this interpretation is incorrect. To check for network failures, we analyzed logs and communication between our RPs and the repositories they queried, confirming their connection. We also downloaded the VRP cache of every RP and analyzed the contents with CURE. Additionally, to mitigate inconsistencies due to network errors or repository states, we started the execution of the RPs simultaneously on a well-connected network. The validation of the RP outputs resulted in the following number of VRP entries on 27th of June 2023: 
\vspace{-5pt}

{\scriptsize{
\begin{verbatim}
Routinator: 441,770 VRPs   |  Fort:        435,002 VRPs
OctoRPKI:   434,074 VRPs   |  rpki-client: 441,777 VRPs
\end{verbatim}
}}
\vspace{-5pt}

The different numbers of VRP entries indicate that processing inconsistencies are prevalent in the RPKI ecosystem. We pin-point the cause of inconsistencies by running all real-world objects against the RPs using the standalone mode of CURE. This allows us to find individual objects for which RPs reach differing conclusions on their validity. However, running real-world objects against the RPs with CURE introduced an additional challenge. These objects are signed by an external entity with an unknown private key, and contain fields indicating specifics of their publication, like domain name, parent authority, or storage location. Thus running these objects without modification in the CURE setup would lead to validation failures of all objects, not allowing any inferences on discrepancies between RPs. To circumvent the problem, we enabled CURE to dynamically adapt to the objects, e.g. creating parent authorities with appropriate names or locations, and signing objects with a CURE private key. With this setup, we found that two inconsistencies that CURE discovered during fuzzing are already affecting validation results of real-world RPKI deployments.
The first processing inconsistency relates to the processing of prefix lengths. We observe that OctoRPKI discards prefixes with a length larger than the maximum allowed prefix length in BGP, /24 for IPv4 and /48 for IPv6. This leads to OctoRPI discarding 1744 prefixes of current RPKI objects.\\
\indent The second inconsistency found by CURE relates to the issuer name in the EE-certificate of ROAs. Certificates that contain issuer names which are not of the type \textit{CommonName} or \textit{SerialNumber} are discarded by Fort. We observe that a subset of objects published in the RPKI use an additional optional issuer name, using the type \textit{OrganisationName}. The majority of these objects are issued by Amazon. Not accepting the \textit{OrganisationName} attribute leads to Fort discarding 6405 IP prefixes, opening them up to prefix hijacks in all systems using Fort. The victims of this problem do not have any means of noticing the missing prefixes without specifically testing if these objects are accepted by Fort. This problem results from the vagueness of the RFC, which leads to unprotected resources. This is especially severe for a cloud provider like Amazon since many systems might have outsourced services into the cloud, thus increasing the scope of the hijack and making the outcome extremely critical.\\
\indent These differences are problematic, as objects, deemed valid by one RP, are not accepted by another, expose the protected prefixes to hijacks. The resource owner that issues the ROA has no indication which RPs accept the published objects, and it is thus likely that many resource owners are not aware that their resources are not protected in networks with certain RPs.\\
\indent {\bf Affected prefixes.} We analyze the inconsistencies in the VRP caches and discovered major companies among the affected victims: Amazon (AS14618 \& AS16509), Microsoft (AS8075), IaaS provider DediPath (AS35913), cybersecurity company Path Networks (AS396998) and global Internet provider Hurricane (AS6939). These entities have several prefixes unprotected by Fort and OctoRPKI i.e., 13\% of all RPs do not propagate many of their ROAs to routers. We observe a total discrepancy of almost 6K prefixes not covered by OctoRPKI/Fort but present in the Routinator/rpki-client VRP cache. We discover 229 ASes do not have all their prefixes covered in Fort, compared to full coverage in rpki-client and Routinator. The number goes up to 526 ASes for OctoRPKI. Despite offering the best VRP coverage on the Internet, we find inconsistent coverage between Routinator and rpki-client as well. Namely, Routinator provides coverage for all resources of large providers like the Saudi Telecom Company and TDC Holdings (formerly, Tele Denmark Communications) while rpki-client does not accept major /12-14 IPv4 and /46-48 IPv6 prefixes from them.

\section{\hspace{3mm} Conclusions}\label{sc:conclusions}

RPKI is used as production grade software in already 37.8\% of the networks to authenticate routes in BGP announcements. Vulnerabilities and bugs can be detrimental to security and correctness of RPKI validation. However, fuzzing RP software implementations is hard. RPs are complex and require a multitude of auxiliary objects and cryptographic computations, they support 9 different data types, each with its own processing logic. In this work we explain how to overcome the identified obstacles and develop a novel tool for fuzzing RPs. Thanks to CURE, we discover vulnerabilities and RFC inconsistencies in all popular RP implementations. We identify 17 DoS-inducing crashes, 1 path traversal vulnerability and at least 7 critical RFC inconsistencies. We show how these flaws can be exploited for attacks and we find discrepancies in RPKI validation even under common network conditions.\\
\indent One of the main factors causing the flaws is that RPKI standards provide a general overview of RP functionalities but leave the corner cases and non-critical field interpretations up to the developers. The standards typically need to offer some flexibility of implementation to allow for competition among products meeting the standard. On the other hand, this freedom becomes a problem when non-critical corner cases and permissive parsings can be exploited for attacks or create inconsistent validation results in different RP implementations. It is therefore important to avoid standards that wind up allowing implementations that defeat the purpose of the standard. 
The parsing and validation inconsistencies create a twofold problem: PP software does not have a strict schema to guarantee every input is parsed correctly by every RP in use, and conversely, some ROAs will not be correctly propagated throughout the Internet due to different parsing preferences by RPs, despite being legitimately issued. This is a silent downgrade of RPKI for many legitimate ROA issuers who believe to be protected across the entire Internet, when in fact many providers discard their ROAs for no good reason. Our work shows that the RPKI software has not matured for production deployments. We hope that our work will inspire the community to develop rigorous requirements on the canonical behavior of RPKI validation. We make our fuzzing system and the CURE tool public to enable developers to improve the quality of their RPKI software. The code can be downloaded at \path{https://github.com/rp-cure/rp-cure}.

\section*{Acknowledgements}
We thank our students Baran Akcin and David Breiding from the Cybersecurity-Seminar at Goethe-Universität for discovering the FORT vulnerability. We are grateful to Christian Rossow and Amir Herzberg for their comments on our research. This work has been co-funded by the German Federal Ministry of Education and Research and the Hessen State Ministry for Higher Education, Research and Arts within their joint support of the National Research Center for Applied Cybersecurity ATHENE and by the Deutsche Forschungsgemeinschaft (DFG, German Research Foundation) SFB~1119.

{
\footnotesize
\bibliographystyle{IEEEtran}
\bibliography{main.bib,ref,bib,sec, fuzz}

\begin{thebibliography}{10}
\providecommand{\url}[1]{#1}
\csname url@samestyle\endcsname
\providecommand{\newblock}{\relax}
\providecommand{\bibinfo}[2]{#2}
\providecommand{\BIBentrySTDinterwordspacing}{\spaceskip=0pt\relax}
\providecommand{\BIBentryALTinterwordstretchfactor}{4}
\providecommand{\BIBentryALTinterwordspacing}{\spaceskip=\fontdimen2\font plus
\BIBentryALTinterwordstretchfactor\fontdimen3\font minus \fontdimen4\font\relax}
\providecommand{\BIBforeignlanguage}[2]{{%
\expandafter\ifx\csname l@#1\endcsname\relax
\typeout{** WARNING: IEEEtran.bst: No hyphenation pattern has been}%
\typeout{** loaded for the language `#1'. Using the pattern for}%
\typeout{** the default language instead.}%
\else
\language=\csname l@#1\endcsname
\fi
#2}}
\providecommand{\BIBdecl}{\relax}
\BIBdecl

\bibitem{bellovin1989security}
S.~M. Bellovin, ``Security problems in the tcp/ip protocol suite,'' \emph{ACM SIGCOMM Computer Communication Review}, vol.~19, no.~2, pp. 32--48, 1989.

\bibitem{china:telecom}
Arstechnica, ``{BGP event sends European mobile traffic through China Telecom for 2 hours},'' \\\url{https://arstechnica.com/informationtechnology/2019/06/bgp-mishap-sends-europeanmobile-traffic-through-china-telecom\\ -for-2-hours}, 2019.

\bibitem{ballani2007study}
H.~Ballani, P.~Francis, and X.~Zhang, ``{A Study of Prefix Hijacking and Interception in the Internet},'' in \emph{ACM SIGCOMM Computer Communication Review}, vol.~37.\hskip 1em plus 0.5em minus 0.4em\relax ACM, 2007, pp. 265--276.

\bibitem{fb:out}
S.~Janardhan, ``{More details about the October 4 outage},'' \\\url{https://engineering.fb.com/2021/10/05/networking-traffic/outage-details/}, 2021.

\bibitem{u:tube}
R.~NCC, ``{YouTube Hijacking: A RIPE NCC RIS case study},'' 2008.

\bibitem{mitm:threat}
Renesys, ``{The New Threat: Targeted Internet Traffic Misdirection},'' \\\url{http://www.renesys.com/2013/11/mitm-internet-hijacking/}, 2013.

\bibitem{indosat:hijack}
A.~Toonk, ``{Hijack Event Today by Indosat},'' \\\url{http://www.bgpmon.net/hijack-event-today-by-indosat/}, 2014.

\bibitem{turkey:hijack}
{A. Toonk}, ``{Turkey Hijacking IP Addresses for Popular Global DNSProviders},'' \\\url{https://www.bgpmon.net/turkey-hijacking-ip-addresses-for-popular-\\ -global-dns-providers/}, 2014.

\bibitem{vervier2015mind}
P.-A. Vervier, O.~Thonnard, and M.~Dacier, ``{Mind Your Blocks: On the Stealthiness of Malicious BGP Hijacks},'' in \emph{NDSS}, 2015.

\bibitem{lepinski2017rfc}
M.~Lepinski and K.~Sriram, ``Rfc 8205: Bgpsec protocol specification,'' 2017.

\bibitem{aspa}
A.~Azimov, E.~Bogomazov, R.~Bush, K.~Patel, and J.~Snijders, ``{Verification of AS\_PATH Using the Resource Certificate Public Key Infrastructure and Autonomous System Provider Authorization},'' November, 2020, \url{https://datatracker.ietf.org/doc/html/draft-ietf-sidrops-aspa-verification-06}.

\bibitem{amir:ndss:2024}
C.~Morris, A.~Herzberg, B.~Wang, and S.~Secondo, ``Bgp-isec: Improved security of internet routing against post-rov attacks,'' in \emph{NDSS}, 2024.

\bibitem{DBLP:conf/uss/HlavacekSVW23}
T.~Hlavacek, H.~Schulmann, N.~Vogel, and M.~Waidner, ``Keep your friends close, but your routeservers closer: Insights into {RPKI} validation in the internet,'' in \emph{32nd {USENIX} Security Symposium, {USENIX} Security 2023, Anaheim, CA, USA, August 9-11, 2023}, J.~A. Calandrino and C.~Troncoso, Eds.\hskip 1em plus 0.5em minus 0.4em\relax {USENIX} Association, 2023, pp. 4841--4858.

\bibitem{DBLP:conf/ccs/HlavacekJMSW22}
T.~Hlavacek, P.~Jeitner, D.~Mirdita, H.~Shulman, and M.~Waidner, ``Behind the scenes of {RPKI},'' in \emph{Proceedings of the 2022 {ACM} {SIGSAC} Conference on Computer and Communications Security, {CCS} 2022, Los Angeles, CA, USA, November 7-11, 2022}.\hskip 1em plus 0.5em minus 0.4em\relax {ACM}, 2022, pp. 1413--1426.

\bibitem{shulman2022poster}
H.~Shulman, N.~Vogel, and M.~Waidner, ``Poster: Insights into global deployment of rpki validation,'' in \emph{Proceedings of the 2022 ACM SIGSAC Conference on Computer and Communications Security}, 2022, pp. 3467--3469.

\bibitem{rovista}
\BIBentryALTinterwordspacing
T.~Chung and W.~Li, ``Rovista,'' 2023. [Online]. Available: \url{https://rovista.netsecurelab.org/}
\BIBentrySTDinterwordspacing

\bibitem{apnicROV}
\BIBentryALTinterwordspacing
APNIC, ``Rov measurement,'' 2023, accessed: 28.06.2023. [Online]. Available: \url{https://stats.labs.apnic.net/rpki}
\BIBentrySTDinterwordspacing

\bibitem{usenix-stalloris-21}
\BIBentryALTinterwordspacing
``{Stalloris: {RPKI} Downgrade Attack},'' in \emph{31st USENIX Security Symposium (USENIX Security 22)}.\hskip 1em plus 0.5em minus 0.4em\relax Boston, MA: USENIX Association, Aug. 2022. [Online]. Available: \url{https://www.usenix.org/conference/usenixsecurity22/presentation/hlavacek}
\BIBentrySTDinterwordspacing

\bibitem{DBLP:conf/sigcomm/HlavacekJMSW23}
T.~Hlavacek, P.~Jeitner, D.~Mirdita, H.~Schulmann, and M.~Waidner, ``Beyond limits: How to disable validators in secure networks,'' in \emph{Proceedings of the {ACM} {SIGCOMM} 2023 Conference, {ACM} {SIGCOMM} 2023, New York, NY, USA, 10-14 September 2023}, H.~Schulzrinne, V.~Misra, E.~Kohler, and D.~A. Maltz, Eds.\hskip 1em plus 0.5em minus 0.4em\relax {ACM}, 2023, pp. 950--966.

\bibitem{mirdita2022poster}
D.~Mirdita, H.~Shulman, and M.~Waidner, ``Poster: Rpki kill switch,'' in \emph{Proceedings of the 2022 ACM SIGSAC Conference on Computer and Communications Security}, 2022, pp. 3423--3425.

\bibitem{koenvanhove}
\BIBentryALTinterwordspacing
K.~van Hove, J.~van~der Ham, and R.~van Rijswijk{-}Deij, ``Rpkiller: Threat analysis from an {RPKI} relying party perspective,'' \emph{CoRR}, vol. abs/2203.00993, 2022. [Online]. Available: \url{https://doi.org/10.48550/arXiv.2203.00993}
\BIBentrySTDinterwordspacing

\bibitem{liang2018fuzzing}
H.~Liang, X.~Pei, X.~Jia, W.~Shen, and J.~Zhang, ``Fuzzing: State of the art,'' \emph{IEEE Transactions on Reliability}, vol.~67, no.~3, pp. 1199--1218, 2018.

\bibitem{ba2022stateful}
J.~Ba, M.~B{\"o}hme, Z.~Mirzamomen, and A.~Roychoudhury, ``Stateful greybox fuzzing,'' in \emph{31st USENIX Security Symposium (USENIX Security 22)}, 2022, pp. 3255--3272.

\bibitem{mckeeman1998differential}
W.~M. McKeeman, ``Differential testing for software,'' \emph{Digital Technical Journal}, vol.~10, no.~1, pp. 100--107, 1998.

\bibitem{durumeric2013zmap}
Z.~Durumeric, E.~Wustrow, and J.~A. Halderman, ``Zmap: Fast internet-wide scanning and its security applications.'' in \emph{Usenix Security}, vol. 2013, 2013.

\bibitem{partridge2016ethical}
C.~Partridge and M.~Allman, ``Ethical considerations in network measurement papers,'' \emph{Communications of the ACM}, vol.~59, no.~10, pp. 58--64, 2016.

\bibitem{DBLP:conf/uss/KruppGR22}
J.~Krupp, I.~Grishchenko, and C.~Rossow, ``Ampfuzz: Fuzzing for amplification ddos vulnerabilities,'' in \emph{31st {USENIX} Security Symposium, {USENIX} Security 2022, Boston, MA, USA, August 10-12, 2022}, K.~R.~B. Butler and K.~Thomas, Eds.\hskip 1em plus 0.5em minus 0.4em\relax {USENIX} Association, 2022, pp. 1043--1060.

\bibitem{schumilo2017kafl}
S.~Schumilo, C.~Aschermann, R.~Gawlik, S.~Schinzel, and T.~Holz, ``kafl: Hardware-assisted feedback fuzzing for os kernels.'' in \emph{USENIX Security Symposium}, 2017, pp. 167--182.

\bibitem{de2015protocol}
J.~De~Ruiter and E.~Poll, ``Protocol state fuzzing of $\{$TLS$\}$ implementations,'' in \emph{24th $\{$USENIX$\}$ Security Symposium ($\{$USENIX$\}$ Security 15)}, 2015, pp. 193--206.

\bibitem{DBLP:conf/uss/KandeCPJSTR22}
R.~Kande, A.~Crump, G.~Persyn, P.~Jauernig, A.~Sadeghi, A.~Tyagi, and J.~Rajendran, ``Thehuzz: Instruction fuzzing of processors using golden-reference models for finding software-exploitable vulnerabilities,'' in \emph{31st {USENIX} Security Symposium, {USENIX} Security 2022, Boston, MA, USA, August 10-12, 2022}, K.~R.~B. Butler and K.~Thomas, Eds.\hskip 1em plus 0.5em minus 0.4em\relax {USENIX} Association, 2022, pp. 3219--3236.

\bibitem{eddington2011peach}
M.~Eddington, ``Peach fuzzing platform,'' \emph{Peach Fuzzer}, vol.~34, pp. 32--43, 2011.

\bibitem{AFLplusplus-Woot20}
A.~Fioraldi, D.~Maier, H.~Ei{\ss}feldt, and M.~Heuse, ``{AFL++}: Combining incremental steps of fuzzing research,'' in \emph{14th {USENIX} Workshop on Offensive Technologies ({WOOT} 20)}.\hskip 1em plus 0.5em minus 0.4em\relax {USENIX} Association, Aug. 2020.

\bibitem{serebryany2017oss}
K.~Serebryany, ``Oss-fuzz-google’s continuous fuzzing service for open source software,'' in \emph{USENIX Security symposium}.\hskip 1em plus 0.5em minus 0.4em\relax USENIX Association, 2017.

\bibitem{kakarla2022scale}
S.~K.~R. Kakarla, R.~Beckett, T.~Millstein, and G.~Varghese, ``$\{$SCALE$\}$: Automatically finding $\{$RFC$\}$ compliance bugs in $\{$DNS$\}$ nameservers,'' in \emph{19th USENIX Symposium on Networked Systems Design and Implementation (NSDI 22)}, 2022, pp. 307--323.

\bibitem{wang2013rpfuzzer}
Z.~Wang, Y.~Zhang, and Q.~Liu, ``Rpfuzzer: A framework for discovering router protocols vulnerabilities based on fuzzing.'' \emph{KSII Transactions on Internet \& Information Systems}, vol.~7, no.~8, 2013.

\bibitem{godefroid2008automated}
P.~Godefroid, M.~Y. Levin, D.~A. Molnar \emph{et~al.}, ``Automated whitebox fuzz testing.'' in \emph{NDSS}, vol.~8, 2008, pp. 151--166.

\bibitem{godefroid2008grammar}
P.~Godefroid, A.~Kiezun, and M.~Y. Levin, ``Grammar-based whitebox fuzzing,'' in \emph{Proceedings of the 29th ACM SIGPLAN conference on programming language design and implementation}, 2008, pp. 206--215.

\bibitem{bohme2016coverage}
M.~B{\"o}hme, V.-T. Pham, and A.~Roychoudhury, ``Coverage-based greybox fuzzing as markov chain,'' in \emph{Proceedings of the 2016 ACM SIGSAC Conference on Computer and Communications Security}, 2016, pp. 1032--1043.

\bibitem{yan2020pathafl}
S.~Yan, C.~Wu, H.~Li, W.~Shao, and C.~Jia, ``Pathafl: Path-coverage assisted fuzzing,'' in \emph{Proceedings of the 15th ACM Asia Conference on Computer and Communications Security}, 2020, pp. 598--609.

\bibitem{afl}
\BIBentryALTinterwordspacing
Google, ``american fuzzy lop,'' 2023. [Online]. Available: \url{https://github.com/google/AFL}
\BIBentrySTDinterwordspacing

\bibitem{brubaker2014using}
C.~Brubaker, S.~Jana, B.~Ray, S.~Khurshid, and V.~Shmatikov, ``Using frankencerts for automated adversarial testing of certificate validation in ssl/tls implementations,'' in \emph{2014 IEEE Symposium on Security and Privacy}.\hskip 1em plus 0.5em minus 0.4em\relax IEEE, 2014, pp. 114--129.

\bibitem{zhou2020web}
X.~Zhou and B.~Wu, ``Web application vulnerability fuzzing based on improved genetic algorithm,'' in \emph{2020 IEEE 4th Information Technology, Networking, Electronic and Automation Control Conference (ITNEC)}, vol.~1.\hskip 1em plus 0.5em minus 0.4em\relax IEEE, 2020, pp. 977--981.

\bibitem{wang2019superion}
J.~Wang, B.~Chen, L.~Wei, and Y.~Liu, ``Superion: Grammar-aware greybox fuzzing,'' in \emph{2019 IEEE/ACM 41st International Conference on Software Engineering (ICSE)}.\hskip 1em plus 0.5em minus 0.4em\relax IEEE, 2019, pp. 724--735.

\bibitem{torres2020nfdfuzz}
G.~Torres, D.~Pesavento, J.~Shi, and L.~Benmohamed, ``Nfdfuzz: A stateful structure-aware fuzzer for named data networking,'' in \emph{Proceedings of the 7th ACM Conference on Information-Centric Networking}, 2020, pp. 169--171.

\bibitem{kim2022efficient}
H.~Kim, Y.~Jeong, W.~Choi, D.~H. Lee, and H.~J. Jo, ``Efficient ecu analysis technology through structure-aware can fuzzing,'' \emph{IEEE Access}, vol.~10, pp. 23\,259--23\,271, 2022.

\bibitem{atheris}
\BIBentryALTinterwordspacing
Google, ``Atheris: A coverage-guided, native python fuzzer,'' 2023. [Online]. Available: \url{https://github.com/google/atheris}
\BIBentrySTDinterwordspacing

\bibitem{cisco:rpki}
Cisco, ``{ Routing Configuration Guide for Cisco ASR 9000 Series Routers, IOS XR Release 6.2.x },'' \\\url{https://www.cisco.com/c/en/us/td/docs/routers/asr9000/software/asr9k-r6-2/routing/configuration/guide/b-routing-cg-asr9000-62x/b-routing-cg-asr9000-62x_chapter_010.html}, 2023.

\bibitem{juniper:rpki}
J.~Networks, ``{Junos OS: BGP User Guide},'' \\\url{https://www.juniper.net/documentation/us/en/software/junos/bgp/topics/ref/statement/session-edit-routing-options-validation.html}, 2023.

\bibitem{frr:rpki}
FRR, ``{FRRouting},'' \\\url{https://docs.frrouting.org/en/latest/bgp.html#configuring-rpki-rtr-cache-servers}, 2023.

\bibitem{nist}
N.~R. Monitor, ``{NIST RPKI Monitor},'' \\\url{https://rpki-monitor.antd.nist.gov/}, 2023.

\bibitem{friedemann2022assessing}
P.~H. Friedemann, N.~Rodday, and G.~D. Rodosek, ``Assessing the rpki validator ecosystem,'' in \emph{2022 Thirteenth International Conference on Ubiquitous and Future Networks (ICUFN)}.\hskip 1em plus 0.5em minus 0.4em\relax IEEE, 2022, pp. 295--300.

\end{thebibliography}
}

\appendices
\section{Detailed Overview of DoS Errors}\label{bugs}

\begin{table*}[t!]
\scriptsize
\renewcommand{\arraystretch}{0.5}
    \centering
  \begin{tabular}{l|c|c|c|c|c}
 Relying Party & ROA & MFT & CRL & CERT & Other \\
 \hline
 Routinator &\makecell{2\\ \{bcder/source.rs:442, \\ rpki-rs/ipres.rs:1237\}} &\makecell{3\\ \{bcder/oid.rs:115, \\ rpki-rs/manifest.rs:372, \\ bcder/tag.rs:429\}}& - &\makecell{4\\ \{bcder/source.rs:547-661, \\ bcder/octet.rs:598, \\ bcder/oid.rs:373\}}& \makecell{2\\ \{rrdp.rs:1974, rpki-rs/uri.rs:107\}}\\
 Fort & - & - & - & - &\makecell{1 \\\{pdu.h:134 \}} \\
 OctoRPKI & - &\makecell{3 \\ \{ber.go:146-152-194\}}&\makecell{2 \\ \{pki.go:146-359\}}& - &\makecell{1 \\ \{octorpki.go:1318\}} \\
 rpki-client & - & - & - & - & - \\
  \end{tabular}
  \caption{Software issues triggering crashes and path traversal in RP implementations.}
  \label{tab:2}
  \vspace{-20pt}
\end{table*}

\subsection{Routinator}

\begin{table}
\scriptsize
\renewcommand{\arraystretch}{0.8}
\centering
\begin{tabular}{l|c|c}
 & crashID & trigger:line \\ \hline
MFT \#1 & 8MhhkLnKUyVS &   bcder/oid.rs:115\\ 
ROA \#2& 6mFeYdrlmPIh &  bcder/source.rs:442 \\ 
ROA \#3 & ZxCVpV04nURw & rpki-rs/ipres.rs:1237 \\ 
MFT \#4 &  27rOGfrl8PuH & bcder/tag.rs:429 \\ 
CER \#5 & fassdfaewwe & bcder/source.rs:547 \\ 
CER \#6 & fassdfaewwe\_2 & bcder/source.rs:661 \\ 
CER \#7 & fassdfaewwe\_3 & bcder/octet.rs:598 \\ 
CER \#8 & fassdfaewwe\_4  &  bcder/oid.rs:373 \\ 
MFT \#9 & XahhkLnKUyVS &  rpki-rs/manifest.rs:372  \\ 

\end{tabular}
\end{table}

\begin{enumerate}

    \item \textbf{MFT\#1} is triggered through manifest processing when content is truncated and the manifest oid is not correctly encoded. Error is thrown in the external library \textit{bcder}.
\begin{lstlisting}[linewidth=\columnwidth,breaklines=true,basicstyle=\small]
if &content.slice()[..len] == 
    self.0.as_ref() {
        content.skip_all()?;
        Ok(())}
\end{lstlisting}

    \item  \textbf{ROA\#2} applies to ROA processing. It is triggered by the decoding library \textit{bcder} in \textit{source.rs:442}. The error raises an exception and causes a crash when the given content length is set lower than the length of the actual content. This error is triggered in \textit{engine.rs:792} of Routinator when the erroneous manifest is first getting decoded.

\begin{lstlisting}[linewidth=\columnwidth,breaklines=true,basicstyle=\small]
if let Some(cur) = self.limit {
    match limit {
        Some(limit)=>assert!(limit<=cur),
        None => panic!("relimiting to unlimited"),
    }
}
\end{lstlisting}

    \item \textbf{ROA\#3} is triggered by the \textit{rpki-rs} library, namely the file \textit{ipres.rs} on line 1237. The piece of code that triggered the crash was meant to create the Prefix object for Routinator, however the prefix length should not be larger than 128 bits and in case it is the code panics due to an assertion error. This leads to a crash in Routinator. The following is the function in \textit{rpki-rs} that leads to the crush. 

\begin{lstlisting}[linewidth=\columnwidth,breaklines=true,basicstyle=\small]
pub fn new<A:Into<Addr>>(addr: A,len: u8) 
    -> Self { assert!(len <= 128);
    Prefix {         
        addr: addr.into().to_min(len),            
        len
    }
}
\end{lstlisting}

\item  \textbf{MFT\#4} applies to manifest processing and is triggered by the decoding library \textit{bcder} in \textit{tag.rs:429}. This crash is caused by an uncaught error where length of tag data does not match expectations.
\begin{lstlisting}[linewidth=\columnwidth,breaklines=true,basicstyle=\small]
data[i] = source.slice()[i];
\end{lstlisting}

\item \textbf{CER\#5} This crash is thrown by the  \textit{bcder} library when processing certificates due to an assertion statement. The data malformation is a length non-compliant certificate.
\begin{lstlisting}[linewidth=\columnwidth,breaklines=true,basicstyle=\small]
fn advance(&mut self, len: usize) {
    if let Some(limit) = self.limit {
    assert!(
        len <= limit,
        "advanced past end of limit"
    );
\end{lstlisting}

\item  \textbf{CER\#6} appears when malformed certificates are processed.  The error throws a panic due to abrupt end of data. 
\begin{lstlisting}[linewidth=\columnwidth,breaklines=true,basicstyle=\small]
fn advance(&mut self, len: usize) {
    assert!(
        self.len >= self.pos + len,
        "advanced past the end of data"
    );
    self.pos += len;}
\end{lstlisting}

\item \textbf{CER\#7} is triggered by another external decoding library \textit{bcder/src/decode/source.rs} for malformed certificates. The crash happens due to the failed decoding of octet strings with a misplaces assertion.
\begin{lstlisting}[linewidth=\columnwidth,breaklines=true,basicstyle=\small]
fn advance(&mut self, len: usize) {
assert!(len <= self.current.len());
self.pos = self.pos + len.into();
bytes::Buf::advance(&mut self.current,len)
}
\end{lstlisting}

\item \textbf{CER\#8} is caused by a panic during certificate processing when the Attribute Value of one of the RDNSequence elements does not comply with the expected type (here PrintableString).
\begin{lstlisting}[linewidth=\columnwidth,breaklines=true,basicstyle=\small]
panic!("illegal object identifier 
(last octet has bit 8 set)");
\end{lstlisting}

\item  \textbf{MFT\#9}  is triggered by the \textit{rpki-rs} library during manifest parsing in file \textit{manifest.rs}. The error is thrown when the manifest contains an invalid character.
\begin{lstlisting}[linewidth=\columnwidth,breaklines=true,basicstyle=\small]
fn next(&mut self)->Option<Self::Item>{
    self.0.decode_partial(|cons| {
        FileAndHash::take_opt_from(cons)
    }).unwrap()
}
\end{lstlisting}

\end{enumerate}

\subsection{OctoRPKI}

\begin{table}
\scriptsize
\renewcommand{\arraystretch}{0.8}
\centering
\begin{tabular}{l|c|c}
 & crashID & trigger:line \\ \hline
MFT\#1 & fYjzQloFL7ej & ber.go:146  \\ 
MFT\#2& fYjzQloFL7ej\_2 & ber.go:152  \\ 
MFT\#3 &  tR1izFWr3V8q &  ber.go:194\\ 
CRL\#4 & wmS41yM39Hau & pki.go:359  \\ 
CRL\#5 & wmS41yM39Hau\_2  &  pki.go:146  \\ 

\end{tabular}
\end{table}

\begin{enumerate}
    \item \textbf{MFT\#1} is triggered during the decoding of a manifest from BER to DER format and the length of a field is shorter than expected by the format hence an "index out of range" panic.
\begin{lstlisting}[linewidth=\columnwidth,breaklines=true,basicstyle=\small]
if tag == 0x1F {
    tag = 0
    for ber[offset] >= 0x80 {
        tag=tag*128 + ber[offset]- 0x80
		offset++
    }
    tag=tag*128 + ber[offset] - 0x80
    offset++
}
\end{lstlisting}

    \item \textbf{MFT\#2} is similar in nature to MFT\#1 except for being triggered at another bytetraversing part of the code. Misconfigured MFT content will trigger a variety of bugs in the source code.
    \item \textbf{MFT\#3} is another crash triggered applies during the reading and decoding of a BER encoded manifest. When the content is shorter than expected a runtime index out of error is thrown \textit{panic: runtime error: index out of range [199] with length 199}.
    \item \textbf{CRL\#4} is triggered when a faulty AuthorityKeyIdentifier is supplied in a CRL. The problematic line is 364 in \textit{pki.go}. OctoRPKI will set the valid value to false and it recognizes that the object is not valid because the AKI is incorrect, but it still accesses the \textit{parent.File} value, which leads to a nil-pointer expression thus crashing the client.
\begin{lstlisting}[linewidth=\columnwidth,breaklines=true,basicstyle=\small]
if pkifile.Parent.Parent.Path != 
  res.Parent.File.Path {
    return false, nil, nil, 
    fmt.Errorf("CRL %s does not match 
    with the parent %s", pkifile.Path, 
    pkifile.Parent.Parent.Path)
  }
\end{lstlisting}
    \item \textbf{Crash\#5} is triggered by the \textit{readObject} function in \textit{ber.go}. The function assumes appropriate formatting and the respecting of tagging bytes and has appropriate size expectations, without any checks and catches for noncomforming files. As a result malformations can throw runtime errors and crash the entire application.

\end{enumerate}

\ignore{
\subsection{Fort}

\begin{lstlisting}[linewidth=\columnwidth,breaklines=true,basicstyle=\small]
validation_run_first(void)
{
	int error;
	if (config_get_mode() == SERVER)
		pr_op_warn("First validation cycle has begun, wait until the next notification to connect your router(s)");
	else
		pr_op_warn("The validation has begun.");
	error = vrps_update(NULL);
	if (error)
		return pr_op_err("First validation wasn't successful.");
	if (config_get_mode() == SERVER)
		pr_op_warn("First validation cycle successfully ended, now you can connect your router(s)");
	else
		pr_op_warn("The validation has successfully ended.");
	return 0;
}
\end{lstlisting}
}

\section{Path Traversal - Vulnerable Code}
\begin{figure}[h!]
\begin{lstlisting}
function from_b_https(bytes) -> Result {
    check_uri_ascii(&bytes)?;
    scheme, start = from_prefix(bytes)?;
    if !scheme.is_https() {
        return Err(Error::BadScheme)
    }
    ... // Return Result
 }
\end{lstlisting}
\caption{RRDP URI Parsing - Without Path Traversal check.}
\label{fig:rrdp-pt}
\vspace{-5mm}
\end{figure}

\begin{figure}[h!]
\begin{lstlisting}
 function from_b_rsync(bytes) -> Result{
   check_uri_ascii(&bytes)?;
   if !starts_with(&bytes, b"rsync://"){
       return Err(Error::BadScheme)
   }
   check_path(bytes)?;

   ... // Check URI has at least 3x '/'
   ... // Return Result
 }

\end{lstlisting}
\caption{Rsync URI Parsing - With Path Traversal check.}
\label{fig:rsync-pt}
\vspace{-5mm}
\end{figure}

\begin{figure}[h!]
\begin{lstlisting}
 function check_path(path) -> Result {
    let mut items = path.split('/');
    loop {
        ... // Check if item exists
        if item.is_empty() {
            break
        }
        if item == b".." || item == b"." {
            return Error
        }
    }
    if items.next().is_some() {
        return Error
    }
 }
\end{lstlisting}
\caption{Function to check for path traversals.}
\label{fig:rsync-es}
\vspace{-5mm}
\end{figure}
\section{RFC Inconsistency Detailed Breakdown}\label{rfc-inconsistency}

\textbf{RFC 9286 Section 4} suggests that users match the validity interval between manifest and CRL. However, lack of matching does is not immediate ground for discarding the content of the CA. The paragraph in question: \textit{Each manifest encompasses a CRL, and the nextUpdate field of the manifest SHOULD match that of the CRL's nextUpdate field, as the manifest will be reissued when a new CRL is published}. RPs behave different with regards to this optional suggestion thus leading to processing discrepancies. In our experiment, if we feed RPs with a valid manifest, a valid ROA and an expired CRL, all parties will drop the manifest's content except for OctoRPKI which continues to serve the ROA despite an expired CRL.\\
\indent \textbf{RFC 9286 Section 5.1.2} the RFC specifies the generation format and validation musts for manifests. We observe inconsistencies in paragraph 4. 
The following requirement \textit{"The validity interval of the EE certificate MUST exactly match the thisUpdate and nextUpdate times specified in the manifest's eContent.  (An RP MUST NOT consider misalignment of the validity interval in and of itself to be an error.)"} None of the RPs check or care about the misalignment issue and such instances are ignored.\\
\indent \textbf{According to RFC6487 Section 5} the CRL must fulfill certain conditions to be acceptable, namely \textit{"An RPKI CA MUST include the two extensions, Authority Key Identifier and CRL Number, in every CRL that it issues."} but according to our experiment if the CRL Number extension is missing: OctoRPKI and rpki-client continue to process the CRL thus disregarding the RFC.
Additionally, the RPs behave differently according to the parsing of CRLs even when they are provided. Namely when the \textit{signature} field in the CRL schema is false or missing, OctoRPKI continues to accept the file. The same issue happens when the schema fields \textit{signatureAlgorithm} or \textit{issuer} are missing from the CRL. Additionally the CRL number is expected to have a specific format in Routinator. 
This is a strict rule of formatting that only Routinator follows and is not specifically defined in any RFC, therefore any non-compliant CRL will be discarded. However, all other RPs parse the value as is. The field in question is a non-critical extension but it does lead to Routinator alone discarding entire CRLs when non-compliant. Lastly, lack of CRL in the MFT is accepted by OctoRPKI as a potential scenario and content is processed, other RPs will deny that manifest.\\
\indent \textbf{RFC6482 Section 3}\label{rfc6482} describes the components of the ROA schema and specifically how the field \textit{ipAddrBlocks} should be configured. The RPKI RFC in itself is not very clear on the appropriate formatting of this sequence \textit{"The ipAddrBlocks field encodes the set of IP address prefixes to which the AS is authorized to originate routes. Note that the syntax here is more restrictive than that used in the IP address delegation extension defined in RFC 3779..."} Accordingly, if we look at the core RFC3779 Section 2.2.3.1, the definition of \textit{ipAddrBlocks} is as follows \textit{The IPAddrBlocks type is a SEQUENCE OF IPAddressFamily types.}. Unlike SETs, Sequences do allow for repetition of elements. The RFCs therefore are not strict in the way \textit{ipAddrBlocks} should organized i.e. if the field can contain multiple sequence elements with the same \textit{addressFamily} or if it can only contain one element for each address family. Routinator is very strict in this regard, it will not process ROAs that have multiple entries of the same family within the \textit{ipAddrBlocks}, but all the other RPs will behave differently and accept such ROAs. This behavior is not explicitly defined in the RFC and Routinator being the most used RP is very strict in this regard. This could cause ROA loss if some publication point does not comply with this restrictive not explicit approach. 

\begin{figure}[h!]
    \centering
    \includegraphics[width=0.9\columnwidth]{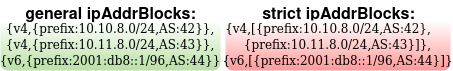}
    \caption{ROA ipAddrBlocks formatting.}
    \label{fig:roa}
    \vspace{-15pt}
\end{figure}

\indent \textbf{RFC6482} has no explicit rules about invalid optional non-critical ROA fields. We observed that when field \textit{MaxLength} is not a parseable integer, all RPs except for OctoRPKI will drop the ROA. This incongruent behavior can lead to unequal ROA distribution across the network and lack of clear guidelines for future publication point software. We also observed RPs processing ROAs differently according to the inheritance status in the EE-certificate of the AS-ID value. Namely, all except rpki-client accept specified inheritance for ROA AS-ID. The RFCs have no specifications in this regard either.\\
\indent \textbf{RFC6487 Section 4}\label{rfc6487} describes the profile and characteristics certificates must display to be acceptable by RPs. Section 4.2 requires for 2 or more serial numbers of certificates to be unique but according to our observations  only Fort and rpki-client adhere to this requirement by denying identical certificates. Routinator and OctoRPKI accept the identical certificate hashes. Section 4.8.4 defines the extension \textit{key\_usage} and \textit{certificate\_resources} as critical and must be present however when these extensions are not sent to critical, again, only Fort and rpki-client enforce the requirement by dropping certificates and the other RPs accept them. \\
\textbf{RFC8182 Section 3.4.3} defines mandatory processing steps for Snapshot. According the the our tests when session\_id in the notification.xml file does not match the session\_id in the snapshot.xml, OctoRPKI is the only RP that does not drop the snapshot file, even though the RFC demands rejection in case of incompatibility. This error can make affected RPs vulnerable to replay attacks.

\vspace{-20pt}
\section{RPKI Objects}\label{ap.sc.objects}
Independent of the mode of operation, the design goal of CURE is the ability to create a valid RPKI repository around arbitrary objects. Understanding how this can be achieved necessitates understanding the role and interaction of all object types in the RPKI.
Generally, objects in the RPKI serve two different purposes, either containing a payload with information for external use, or containing objects to serve the RPKI infrastructure for internal processing in the RPs, ensuring integrity and authenticity. 
Currently, three different objects containing payload and six objects for internal processing are standardized.\\
\indent \textbf{Route Origin Authorizations (ROAs)} are the main payload of the RPKI. Each ROA contains an AS number and a set of IP resources, specified as prefixes or IP ranges. A ROA attests that a given AS number is authorized to announce the IP resources contained in the object over BGP.\\
\indent \textbf{AS Provider Authorizations (ASPAs)} Autonomous System Provider Authorization (ASPA) objects bind a customer AS number and the authorized provider AS number for its announcements. The AS path verification with RPKI checks if the routes received from customers or peers comply with the ASPAs to prevent path manipulation attacks.\\
\indent \textbf{Ghostbuster Records (GBRs)} contain information on contacts responsible for a certain domain. GBRs are currently not commonly used, as we could only observe 3 GBRs currently in use. To distribute these payloads to the BGP routers of ASes, a number of additional objects are standardized that ensure the integrity and authenticity of the payloads.\\
\indent \textbf{X.509 Cryptographic Certificates} are used to ensure authenticity of objects in the RPKI. 
Each certificate contains a private/public RSA key pair that is used to sign child objects, namely either object payloads or other certificates sub-allocating owned resources. The RPKI distinguishes two types of X.509 certificates used. End-Entity (EE) certificates are always used in combination with a single object. They contain a one-time key that is used to sign the digest of the object it authenticates, attesting its validity. EE-Certificates are stored together with the authenticated object inside a single file. While ee-certificates authenticate the validity of individual objects, there do not provide a means to verify the resources claimed in a given payload. Thus the RPKI also utilizes Certificate Authorities (CAs) certificates. Each CA certificate holds information on the resources owned by a specific CA, i.e. owned AS numbers and IP address spaces. A CA is only authorized to sign objects that claim ownership over resources, i.e. IP and ASN ranges, that are stated in the CA certificate. The valid ownership of the resources claimed in the CA certificate is attested by the parent CA that assigned the resources to the CA. This concept of parent ownership leads to the tree-like structure of the RPKI, with each CA sub-allocating parts of their resources to child CAs, who can issue payload objects for their assigned resources or further sub-allocate their resources to their children.\\
 \indent \textbf{Manifests} ensure the integrity of each RPKI repository. Each manifest contains a list of all objects present in a given repository in combination with the hash of the corresponding file. This allows RPs to confirm that no object in the repository has been added, deleted, or manipulated. The manifest is cryptographically signed with an EE-Certificate and needs to be re-issued every time any object in the repository changes. Each RPKI repository should contain a single manifest file.\\
 \indent \textbf{Certificate Revocation Lists (CRLs)} detail which certificates of a specific CA have been revoked. This allows CAs to remove objects even before their validity expired. The CRL is checked by the RPs to conclude which certificates should be discarded, even if their cryptographic validation is successful. Additionally to a manifest, each RPKI repository also needs to contain a single CRL.\\
 \indent The files inside the repository are distributed to the RPs over one of the RPKI communication protocols, rsync or RRDP. While rsync is a simple file synchronization protocol, RRDP defines additional objects that are used to efficiently distribute RPKI objects.\\
 \indent \textbf{Notification.xml} is the entry point to each repository and is always located at a fixed hardcoded location, that is specified in the CA certificate that uses the publication point. The role of the notification file is to notify RPs of the location of the repository content as well as the current repository state. The header of the file contain a session id and a serial number, detailing the current state of the repository. If any object in the repository is changed, the serial number is incremented, informing the RPs that new content should be fetched. The notification file links to the content of the repository by providing the links and digests of the snapshot and delta files.\\
 \indent \textbf{Snapshot.xml} contains the entire repository content. Each entry in the file consists of the filename of the object and its content, as a base64 encoded byte string. The Snapshot is updated each time the repository changes.\\
 \indent \textbf{Delta.xml} files can be used to incrementally update the repository from a previous state. They detail the changes to the repository since the last serial number, i.e. which objects have been added, deleted, and modified. RPs that previously downloaded the snapshot can use the delta files to incrementally update their local cache without downloading the entire snapshot file. If delta processing fails, the RP resorts back to the snapshot file.

\end{document}